\documentclass[nofootinbib,prd,preprintnumbers,superscriptaddress,aps]{revtex4}
\usepackage{amsmath}
\usepackage{amsfonts}
\usepackage{graphicx}
\usepackage[colorlinks=true,linkcolor=blue,citecolor=teal,urlcolor=blue]{hyperref}
\usepackage{xcolor}
\numberwithin{equation}{section}

\usepackage{hyperref}

\usepackage{physics}
\usepackage{bbold}
\usepackage{cases}
\usepackage{braket}
\usepackage[T1]{fontenc}
\usepackage{mathrsfs}
\usepackage{enumerate}
\usepackage{bm}
\usepackage{multirow}
\usepackage{comment}
\usepackage{mathrsfs}

\DeclareMathOperator{\Shi}{Shi}

\makeatother
\begin{document}
	\title{Constraints on Kaniadakis Cosmology from Starobinsky Inflation and Primordial Tensor Perturbations}
	
	\author{Abdelhakim Benkrane}   
	\email{abdelhakim.benkrane@univ-ouargla.dz, hakim9502.benkrane@gmail.com}	
\affiliation{Université Kasdi Merbah Ouargla, Laboratoire LRPPS, Ouargla 30000, Algeria}

\author{Giuseppe Gaetano Luciano}
\email{giuseppegaetano.luciano@udl.cat}
\affiliation{Department of Chemistry, Physics and Environmental and Soil Sciences, Polytechnic School, University of Lleida, Av. Jaume II, 69, 25001 Lleida, Spain
}

\date{\today}

\begin{abstract}
We investigate a generalized entropic cosmology obtained by applying the gravity-thermodynamics conjecture to the Universe horizon using Kaniadakis statistics, namely a relativistic extension of the standard Boltzmann--Gibbs formalism. The resulting deformation of the horizon entropy naturally modifies the Friedmann dynamics and provides a phenomenologically consistent extension of the \(\Lambda\)CDM paradigm. Within this framework, we explore the implications of the modified cosmological dynamics for the physics of the early Universe, focusing in particular on primordial gravitational waves (PGWs) and slow-roll inflation in a Starobinsky-like scenario. We show that the generalized entropic corrections simultaneously affect the evolution of tensor perturbations and the inflationary slow-roll dynamics, inducing characteristic deviations in the PGW spectrum as well as nontrivial corrections to the main inflationary observables. By confronting the theoretical predictions with the latest Planck and BICEP/Keck observations, we derive stringent constraints on the Kaniadakis parameter and assess the observational viability of the model. Our results establish a direct connection between generalized horizon thermodynamics and inflationary cosmology, showing that quantum-statistical modifications of the entropy-area law can propagate into potentially observable signatures in the physics of the early Universe.
\end{abstract}

	\maketitle
    
	\section{Introduction}	
	Modern cosmology has achieved remarkable success in describing the large-scale evolution of the Universe through the standard $\Lambda$CDM model, which is strongly supported by observations of the cosmic microwave background (CMB), large-scale structure and type Ia supernovae. Nevertheless, several fundamental problems remain open, including the origin of dark energy, the mechanism behind the baryon asymmetry of the Universe and the physics driving the inflationary epoch. In particular, cosmological observations indicate that the Universe underwent phases of accelerated expansion both at very early and at late times. 
    
    In order to account for these phenomena, two main theoretical approaches have been extensively investigated in the literature. A first possibility consists in modifying the gravitational sector itself, namely by extending the Einstein--Hilbert action through additional geometrical or higher-curvature contributions. This framework gives rise to several modified theories of gravity, including \(f(R)\) gravity, Gauss--Bonnet extensions, Lovelock theories, Weyl gravity and Galileon models, together with torsional or non-metric formulations such as \(f(T)\) gravity, \(f(T,T_G)\) models and other generalized geometric scenarios \cite{Capozziello:2011et, CANTATA:2021asi}. An alternative approach is based on modifying the matter sector of Einstein equations through the introduction of additional dynamical components, such as scalar inflaton fields or dark-energy fluids, which effectively generate the observed accelerated evolution of the Universe \cite{Olive:1989nu, Bartolo:2004if,Cai:2009zp}.

Beyond the aforementioned approaches to extending the standard cosmological framework, an important line of research is based on the deep connection between gravity and thermodynamics. In a seminal work, Jacobson showed that Einstein’s field equations can be derived from thermodynamical arguments applied to local Rindler horizons~\cite{Jacobson1995}, while later developments by Padmanabhan further supported the interpretation of gravity as an emergent phenomenon associated with spacetime thermodynamics~\cite{Padmanabhan2005}. In this perspective, the Universe can be regarded as a thermodynamical system bounded by the apparent horizon and filled with matter and dark-energy fluids. Within such a framework, the Friedmann equations can be reformulated in terms of the first law of thermodynamics by assuming the holographic Bekenstein--Hawking entropy-area scaling for the cosmological horizon \cite{Bekenstein:1973ur,Bekenstein:1974ax,Hawking:1975vcx}. Conversely, the cosmological dynamics can be reconstructed by applying thermodynamical relations directly on the apparent horizon together with the corresponding entropy law.

Although the Bekenstein--Hawking entropy provides the standard thermodynamical description of gravitational horizons, several arguments suggest that it may represent only an effective semiclassical approximation of a more fundamental underlying relation. In particular, quantum-gravitational effects, non-extensive statistical features and generalized holographic scenarios may lead to deviations from the standard area law, especially in regimes characterized by strong gravitational fields or high-energy cosmological dynamics \cite{Dagotto:1989gp,Carlip:2000nv,Tsallis:2013,Barrow:2020tzx,kaniadakis2001non,renyi1961entropy,Jizba:2024klq,Kaul:2000kf,Bombelli:1986rw,Srednicki:1993im}. Motivated by these considerations, several generalized entropy formalisms have been proposed in the literature as possible extensions of the conventional horizon entropy. Among these, Kaniadakis entropy has attracted considerable attention due to its close connection with relativistic statistical mechanics and its preservation of fundamental symmetries, including Lorentz invariance~\cite{Kaniadakis:2002zz,kaniadakis2001non}. In contrast to other generalized entropic frameworks, the Kaniadakis formalism naturally emerges from relativistic kinetic theory and provides a consistent deformation of standard Boltzmann--Gibbs statistics in high-energy regimes. 

When applied to gravitational and cosmological systems, Kaniadakis entropy modifies the entropy-area relation associated with the cosmological horizon, thereby inducing corrections to the Friedmann equations and to the effective cosmological dynamics. Such corrections can generate an effective dark-energy sector, alter the expansion history of the Universe and modify the evolution of inflationary observables, while still remaining compatible with current observational constraints. In particular, the resulting modifications may affect the slow-roll dynamics, the primordial perturbation spectra and the late-time acceleration behavior, providing a phenomenologically rich framework in which quantum-statistical effects can leave observable cosmological signatures. For these reasons, Kaniadakis entropy has become an increasingly relevant candidate for investigating possible thermodynamical extensions of standard cosmology and gravity \cite{Lymperis:2021qty,Luciano:2022eio,Luciano:2022knb,Hernandez-Almada:2021aiw,Hernandez-Almada:2021rjs}.
    
In this context, the inflationary paradigm \cite{Guth:1980zm, Starobinsky1980, Olive:1989nu,Linde:1983gd,Albrecht:1982wi} and primordial gravitational waves (PGWs) \cite{Starobinsky:1979ty,Grishchuk:1974ny,Rubakov:1982df,Abbott:1984fp,Bernal:2020ywq} provide particularly powerful observational frameworks for testing the cosmological implications of generalized entropy formalisms. Inflation not only offers a successful solution to the horizon and flatness problems of standard cosmology, but also explains the origin of primordial density perturbations through the amplification of quantum fluctuations in the early Universe. At the same time, inflation generically predicts the production of a stochastic background of PGWs, whose imprints may be encoded in the tensor component of cosmological perturbations and in the B-mode polarization of the cosmic microwave background. Since these observables are highly sensitive to the background expansion history and to the underlying gravitational dynamics, they represent an ideal arena for probing possible deviations from standard cosmology induced by generalized entropic corrections.

Among the various inflationary scenarios proposed in the literature, the Starobinsky model remains one of the most successful and observationally favored frameworks, providing predictions in excellent agreement with current cosmological observations~\cite{Starobinsky1980}. In particular, the model naturally predicts values of the scalar spectral index and tensor-to-scalar ratio that are fully compatible with the latest Planck and BICEP/Keck constraints, while supporting a sufficiently long slow-roll phase without requiring significant fine-tuning. Owing to its robustness and predictive power, Starobinsky inflation therefore constitutes an especially suitable framework for investigating the effects of Kaniadakis entropy on the dynamics of the early Universe.

Within this perspective, entropy-induced corrections to the Friedmann equations may modify the slow-roll evolution of the inflaton field and consequently affect key inflationary observables, including the scalar spectral index, its running and the tensor-to-scalar ratio. The study of these effects is therefore particularly relevant, since current and future observations of primordial perturbations and gravitational waves may provide stringent constraints on generalized thermodynamical cosmologies and potentially rule out scenarios predicting deviations incompatible with observational data. At the same time, such observations may offer a unique opportunity to identify possible signatures of quantum-statistical modifications of gravitational dynamics in the early Universe.

Building upon the above discussion, we now investigate how the generalized Kaniadakis entropy framework affects the cosmological dynamics of the early Universe and its associated observational signatures. To this end, we derive the modified Friedmann equations induced by the generalized entropy and analyze the resulting implications for both the background expansion history and the evolution of primordial tensor perturbations. In particular, we study the spectrum of PGWs within the Kaniadakis cosmological scenario and compare the corresponding predictions with those obtained in the standard cosmological framework, emphasizing the characteristic frequency dependence generated by the entropic corrections and their possible observational consequences. 

In addition, we examine the impact of Kaniadakis entropy on slow-roll inflation by considering a Starobinsky-like inflationary potential. Within this setup, we derive the modified slow-roll parameters and evaluate the corresponding inflationary observables, including the scalar spectral index, its running and the tensor-to-scalar ratio. By confronting the theoretical predictions with the latest Planck and BICEP/Keck observational constraints, we obtain stringent bounds on the Kaniadakis parameter and assess the phenomenological viability of the model.

	

The structure of the paper is organized as follows. In Sec.~\ref{KS}, we review the fundamental aspects of Kaniadakis statistics and discuss its holographic application to gravitational systems.  In Sec.~\ref{PGW}, we investigate the spectrum of PGWs within the Kaniadakis cosmological framework, analyzing  how the modified expansion history affects the primordial tensor spectrum. In Sec.~\ref{infl}, we focus on slow-roll inflation by considering a Starobinsky-like  potential. Finally, in Sec.~\ref{Conc}, we summarize the main results and discuss possible future developments. Throughout this work, we adopt natural units such that
\(\hbar=c=k_B=1\).

\section{Foundations of Kaniadakis Statistics}
\label{KS}
In this section, we outline the mathematical and physical principles underlying Kaniadakis statistics. For a comprehensive treatment of the subject, the reader is referred to Ref.~\cite{Kaniadakis:2002zz}.

It is well known that the Maxwell–Boltzmann (MB) distribution constitutes a cornerstone of classical statistical mechanics. Rather than being derived from first principles, it is typically assumed as a fundamental postulate. This observation raises a fundamental question: does the MB distribution also arise when microscopic dynamics are governed by the laws of special relativity? This issue has been investigated in Ref.~\cite{Kaniadakis:2002zz}. Experimental evidence indicates that relativistic cosmic rays exhibit power-law tails in their energy spectra, deviating significantly from the exponential behavior predicted by the MB distribution~\cite{Kaniadakis:2002zz}. Similar features have been observed in various high-energy environments, including plasmas immersed in superthermal radiation fields~\cite{plasma}, nuclear collisions~\cite{NC} and open stellar clusters~\cite{OSC}. These findings suggest the necessity of generalizing the classical Boltzmann–Gibbs–Shannon (BGS) entropy within a relativistic framework.

In Ref.~\cite{Kaniadakis:2002zz}, it has been shown that Lorentz transformations naturally lead to a one-parameter deformation of the Boltzmann–Gibbs entropy, given by
\begin{equation}
\label{KE}
S_{\kappa} = -\sum_{i} n_i \ln_{\kappa} n_i\,,\qquad
\ln_{\kappa} x \equiv \frac{x^{\kappa} - x^{-\kappa}}{2\kappa}.
\end{equation}

The generalized Boltzmann factor corresponding to the $i$-th energy level $E_i$ reads $n_i = \alpha \exp_{\kappa}\!\left[-\beta\left(E_i - \mu\right)\right]$,
where the $\kappa$-exponential function is
\begin{equation}
\label{expk}
\exp_{\kappa}(x) \equiv \left(\sqrt{1 + \kappa^{2}x^{2}} + \kappa x\right)^{1/\kappa}.
\end{equation}
The constants entering the distribution are defined as $\alpha = \left[\frac{1-\kappa}{1+\kappa}\right]^{1/2\kappa}$ and $\frac{1}{\beta} = \sqrt{1-\kappa^{2}}\,T$.

The entropy given in Eq.~\eqref{KE} is known as the \emph{Kaniadakis entropy}. Deviations from the classical BGS framework are governed by the dimensionless parameter $-1 < \kappa < 1$. In the Galilean limit $\kappa \to 0$, standard statistical mechanics is recovered. For convenience, the Kaniadakis entropy can also be expressed as~\cite{Jeans1,Jeans2}
\begin{equation}
\label{KanP}
S_{\kappa} = -\sum_{i=1}^{W}
\frac{P_i^{1+\kappa} - P_i^{1-\kappa}}{2\kappa},
\end{equation}
where $P_i$ denotes the probability of the system being in the $i$-th microstate and $W$ is the total number of accessible configurations.

\subsection{Holographic application of Kaniadakis entropy}

Let us now extend the Kaniadakis paradigm to the black-hole framework \cite{Luciano:2023bai}. This step is particularly useful for holographic considerations and, consequently, for cosmological applications of Kaniadakis entropy. To this end, we assume equiprobable states, $P_i=1/W$, in Eq.~\eqref{KanP}, and use the property that the Boltzmann--Gibbs--Shannon entropy satisfies the relation $S\propto\log W$. Since the Bekenstein--Hawking entropy is given by $S_{BH}=A/(4G)$, it follows that
$W=\exp\left[A/(4G)\right]$.
By substituting this expression into Eq.~\eqref{KanP}, we obtain
\begin{equation}
\label{KenBH}
S_\kappa=\frac{1}{\kappa}\sinh\left(\kappa\hspace{0.4mm}\frac{A}{4G}\right),
\end{equation}
which correctly reproduces the standard Bekenstein--Hawking entropy $S_{BH}$ in the limit $\kappa\rightarrow0$.

Notice that the above function is even, namely $S_\kappa = S_{-\kappa}$. Therefore, one may safely restrict the analysis to the domain $\kappa > 0$. Moreover, since deviations from the Bekenstein--Hawking formula are expected to be small, it is reasonable to perform a perturbative expansion of Eq.~\eqref{KenBH} for $\kappa \ll 1$, neglecting higher-order contributions.

Following Refs.~\cite{Lymperis:2021qty,Drepanou:2021jiv,Luciano:2022knb,Hernandez-Almada:2021rjs}, we now apply the above formalism to the cosmological setting, with particular emphasis on the derivation of modified background evolution equations. In this context, the generalized Kaniadakis entropy introduces corrections to the standard cosmological dynamics, providing a consistent framework to investigate possible deviations from the conventional Friedmann equations and their implications for the evolution of the Universe.

We consider a homogeneous and isotropic Universe described by the Friedmann–Robertson–Walker (FRW) metric
\begin{equation}
ds^{2} = -dt^{2} + a^{2}(t)\left[\frac{dr^{2}}{1-kr^{2}} + r^{2} d\Omega^{2}\right],
\label{metric}
\end{equation}
where $a(t)$ is the scale factor and $k=0,\pm1$ corresponds to flat, closed, and open spatial geometries, respectively. The cosmic content is assumed to be a perfect fluid with energy density $\rho$ and pressure $p$.

In the thermodynamic interpretation of gravity \cite{Jacobson:1995ab}, the Universe is treated as a thermodynamical system bounded by the apparent horizon. The radius of the latter is given by \cite{Frolov:2002va,Cai:2005ra,Akbar:2006er,Akbar:2006kj}
\begin{equation}
r_a = \frac{1}{\sqrt{H^2+\frac{k}{a^2}}},
\label{apparent_horizon}
\end{equation}
where $H=\dot{a}/a$ denotes the Hubble parameter.

Associating thermodynamic quantities with the horizon, its temperature is expressed as \cite{Gibbons:1977mu}
\begin{equation}
T = \frac{1}{2\pi r_a},
\label{temperature1}
\end{equation}
while the entropy depends on the underlying gravitational theory. In General Relativity, the entropy is given by the Bekenstein–Hawking relation \cite{Bekenstein:1973ur,Hawking:1975vcx}
\begin{equation}
S_{BH} = \frac{A}{4G},
\label{SBH}
\end{equation}
where $A=4\pi r_a^2$ is the area of the apparent horizon.

The first law of thermodynamics applied to the horizon reads \cite{Jacobson:1995ab,Cai:2005ra}
\begin{equation}
-dE = T\, dS,
\label{first_law}
\end{equation}
where the energy flux crossing the horizon during an infinitesimal interval $dt$ is $\delta Q = -dE = A(\rho+p)Hr_a\,dt$.

Substituting Eqs.~(\ref{temperature1}) and (\ref{SBH}) into Eq.~(\ref{first_law}), one obtains
\begin{equation}
-4\pi G(\rho+p)=\dot{H}-\frac{k}{a^2}.
\label{raychaudhuri_standard}
\end{equation}

Combining this relation with the conservation equation of the matter fluid
\begin{equation}
\label{conservation}
\dot{\rho}+3H(\rho+p)=0\,,
\end{equation} 
and integrating, we recover the standard Friedmann equation,
\begin{equation}
\frac{8\pi G}{3}\rho=H^2+\frac{k}{a^2}-\frac{\Lambda}{3},
\label{friedmann_standard}
\end{equation}
where $\Lambda$ emerges as an integration constant. This derivation highlights the profound connection between gravitational dynamics and thermodynamics \cite{Jacobson:1995ab,Padmanabhan:2003gd,Padmanabhan:2009vy}. Furthermore, 
in what follows,  we will assume a spatially flat Universe (i.e., $k=0$). Indeed, high-precision cosmological data, including measurements of the cosmic microwave background combined with baryon acoustic oscillations, indicate that the curvature parameter is consistent with zero within very tight bounds~\cite{Planck:2018vyg,Efstathiou:2020wem}.

The above procedure can be generalized by replacing the Bekenstein–Hawking entropy with a generalized entropy expression. In the present framework, the adoption of Kaniadakis entropy \eqref{KenBH} modifies the cosmological dynamics and leads to corrected Friedmann equations. Applying the first law with the generalized entropy, straightforward calculations lead to \cite{Lymperis:2021qty,Drepanou:2021jiv,Luciano:2022knb,Hernandez-Almada:2021rjs}
\begin{equation}
	-4\pi G(\rho+p)=\dot{H}\cosh\left(\dfrac{\kappa \pi}{H^{2}G}\right),
\label{modified_raychaudhuri}
\end{equation}
where $\kappa$ is the dimensionless Kaniadakis parameter.

Using the conservation equation (\ref{conservation}) and integrating, the modified Friedmann equation  becomes
\begin{equation}
\dfrac{8\pi G}{3}\rho=-\dfrac{\kappa \pi}{G}\Shi\left(\dfrac{\kappa \pi}{H^{2}G}\right)+H^{2} \cosh\left(\dfrac{\kappa \pi}{H^{2}G}\right)-\dfrac{\Lambda}{3}\,,
\label{modified_friedmann_flat}
\end{equation}
where $\mathrm{Shi}(x)$ denotes the hyperbolic sine integral,
\begin{equation}
\mathrm{Shi}(x)=\int_{0}^{x}\frac{\sinh t}{t}\,dt.
\end{equation}    
Equation~(\ref{modified_friedmann_flat}), together with the corresponding dynamical relation,
represents the generalized Friedmann equations arising from the adoption of Kaniadakis entropy.
In the limit $\kappa\rightarrow0$, the standard cosmological equations of General Relativity are recovered.

Since Eq.~(\ref{modified_friedmann_flat}) involves transcendental functions, it cannot be solved analytically for the Hubble parameter $H$ in closed form. In order to make further analytical progress, we consider the regime in which the argument of the hyperbolic functions remains small, namely
$\frac{\kappa\pi}{G H^2}\ll 1$.\footnote{
Defining the dimensionless quantity $\beta \equiv \kappa\pi/(G H^2)$, the perturbative expansion of Eq.~\eqref{modified_friedmann_flat} requires $\beta \ll 1$. Since $H(z)$ increases with redshift throughout the cosmic evolution, $\beta$ attains its maximum value at the present epoch, where $H=H_0$. Therefore, the sufficient condition $\beta_0 \equiv \kappa\pi/(G H_0^2)\ll1$ guarantees the validity of the perturbative expansion throughout the entire cosmological history. In Planck units this corresponds to $\kappa \lesssim \mathcal{O}(10^{-123})$~\cite{Hernandez-Almada:2021aiw}.} In this limit, the generalized Kaniadakis entropy represents a small deviation from the standard Bekenstein--Hawking entropy, allowing for a perturbative treatment of the modified cosmological equations. Retaining the leading nontrivial correction, Eq.~(\ref{modified_friedmann_flat}) reduces to
\begin{equation}
\frac{8\pi G}{3}\rho
=
H^2-\frac{\kappa^2\pi^2}{2G^2H^2}-\frac{\Lambda}{3}\,.
\label{modified2}
\end{equation}

Equation~(\ref{modified2}) is quartic in $H$ and admits four solutions. However, only one corresponds to a physically meaningful expanding Universe, namely the positive real branch, which continuously reduces to the standard Friedmann solution in the limit $\kappa \to 0$. Expanding around the standard GR branch and retaining only the leading nontrivial correction in $\kappa$, the physically relevant positive solution becomes 
\begin{equation}
H(t)=
\sqrt{\frac{8\pi G}{3}\rho(t)+\frac{\Lambda}{3}}
+
\frac{\kappa^{2}\pi^{2}}{
4G^{2}\left[\dfrac{8\pi G}{3}\rho(t)+\dfrac{\Lambda}{3}\right]^{3/2}
}\,.
\label{ModHubRate}
\end{equation}
Thus, we observe once again that the standard background evolution given by Eq.~(\ref{friedmann_standard}) is recovered in the limit $\kappa \to 0$.

For the purposes of the following analysis, it is convenient to express the total energy density as a function of the redshift $z$. Assuming the usual decomposition into non-relativistic matter, radiation, spatial curvature and dark-energy components, one can write
\begin{equation}
\rho(z)=\rho_{m,0}(1+z)^3+\rho_{r,0}(1+z)^4+\rho_{k,0}(1+z)^2+\rho_{\Lambda,0}\,,
\label{rho_z}
\end{equation}
where $\rho_{m,0}$, $\rho_{r,0}$, $\rho_{k,0}$, and $\rho_{\Lambda,0}$ denote the corresponding present-day energy densities. 

Equivalently, in terms of the present critical density $\rho_{c,0}=3H_0^2/(8\pi G)$ and the density parameters $\Omega_{i,0}=\rho_{i,0}/\rho_{c,0}$, this relation becomes
\begin{equation}
\rho(z)=\rho_{c,0}
\left[
\Omega_{m,0}(1+z)^3
+\Omega_{r,0}(1+z)^4
+\Omega_{k,0}(1+z)^2
+\Omega_{\Lambda,0}
\right] .
\label{rho_z_omega}
\end{equation}
For a spatially flat background, the curvature term is absent, $\Omega_{k,0}=0$.

\section{PGW Spectrum in Kaniadakis Cosmology}
\label{PGW}

Primordial Gravitational Waves (PGWs) are expected to encode information about quantum fluctuations and possible phase transitions that occurred during the inflationary stage of the early Universe. Their detection would be highly significant, as it could provide a rare observational window into epochs preceding Big Bang Nucleosynthesis (BBN). This includes important stages such as reheating, the quark--hadron transition, and potential non-standard periods in which the cosmic energy density was dominated by matter-like components or kination. Furthermore, since General Relativity is likely to receive quantum corrections at very high energies, the pre-BBN era offers a valuable environment for probing extensions or alternatives to classical gravity in the ultraviolet regime.

In models that go beyond the standard single-field slow-roll inflationary paradigm, gravitational waves may be generated with amplitudes large enough to be detectable at scales smaller than those accessible through observations of the Cosmic Microwave Background (CMB). In this section, we evaluate the spectrum of PGWs within the context of the generalized cosmological framework introduced in Sec.~\ref{KS}, and we contrast these results with those obtained in the standard cosmological model. We also discuss the corresponding observational features and assess their prospects for detection. To carry out this analysis, we adopt the formalism originally developed in Ref.~\cite{Watanabe:2006qe} and further developed in subsequent studies~\cite{Bernal:2020ywq,Barman:2023ktz,Maity:2024cpq,Luciano:2024mcn}.

\subsection{PGWs within the standard cosmological model}

In the linear perturbative framework, gravitational waves are described as small fluctuations of the metric superimposed on a curved spacetime background. We restrict our analysis to tensor modes evolving on a homogeneous, isotropic, and spatially flat Universe. Within this setup, one can consistently impose the conditions $h_{00} = h_{0i} = 0$, which remove temporal and mixed components of the perturbations. In addition, we work in the transverse--traceless (TT) gauge, defined by $\partial_i h_{ij} = 0$ and $h^i_i = 0$, where Latin indices denote spatial coordinates.

Under these assumptions, the first-order evolution equation for tensor perturbations takes the form~\cite{Watanabe:2006qe}
\begin{equation}
\label{hdyn}
\ddot h_{ij} + 3H \dot h_{ij} - \frac{\nabla^2}{a^2} h_{ij} = 16\pi G \hspace{0.3mm} \Pi_{ij}^{TT}\,,
\end{equation}
where $\Pi_{ij}^{TT}$ represents the transverse--traceless component of the anisotropic stress tensor, given by
\begin{equation}
\Pi_{ij} = \frac{T_{ij} - p \hspace{0.3mm} g_{ij}}{a^2}\,,
\label{TPE}
\end{equation}
with $T_{ij}$ denoting the stress--energy tensor, $g_{ij}$ the metric tensor, and $p$ the background pressure.

We emphasize that Eq.~\eqref{hdyn} corresponds to the standard wave equation governing tensor perturbations on a Friedmann--Robertson--Walker (FRW) background, derived within the usual Lagrangian formulation of general relativity. Although its origin does not rely on the thermodynamic considerations underlying the generalized MHR framework, it remains the conventional starting point for analyzing gravitational waves even in non-standard cosmological settings~\cite{Bernal:2020ywq}.

In what follows, we retain Eq.~\eqref{hdyn} in its standard form and explore how the modified background evolution, as dictated by the entropic cosmological model, influences the dynamics and observable properties of the primordial tensor spectrum. In this approach, the thermodynamic framework determines the background expansion history, while tensor perturbations are treated consistently within the standard field-theoretic description. Such a separation allows for a direct comparison with predictions from the standard cosmological model, as well as with the projected sensitivities of upcoming gravitational wave experiments.

A fully self-consistent analysis would ultimately require incorporating perturbations within the entropic framework itself. However, developing such an extension lies beyond the scope of the present work and is left for future study.

Furthermore, it is worth emphasizing that, despite the homogeneous and isotropic nature of the background spacetime, the term $\Pi_{ij}^{TT}$ appearing in Eq.~\eqref{hdyn} captures the possible presence of anisotropic stress perturbations within the cosmic medium. More precisely, it represents the transverse--traceless component of the stress--energy tensor fluctuations, which may originate from a variety of physical sources, including free-streaming relativistic species (such as neutrinos), primordial magnetic fields, or more general imperfect fluid contributions.

In the absence of these sources, tensor perturbations evolve according to the homogeneous part of Eq.~\eqref{hdyn}, corresponding to freely propagating gravitational waves subject only to the damping induced by the expansion of the Universe. On the other hand, when anisotropic stresses are present, they provide a source term that can generate gravitational waves or alter their amplitude and evolution. Consequently, $\Pi_{ij}^{TT}$ plays a key role in shaping the dynamics of tensor modes, as it encodes the influence of microphysical processes and the properties of the cosmic matter content on the gravitational wave background.

Gravitational wave signals are commonly categorized based on their production mechanisms into three broad classes: inflationary, cosmological, and astrophysical sources. The typical frequency of the resulting signals is closely tied to the underlying generation process. In this work, we concentrate on gravitational waves within the frequency range $[10^{-11}, 10^{3}]\,\mathrm{Hz}$, which is particularly significant as it lies within the sensitivity window of both current and forthcoming gravitational wave detectors.

To solve Eq.~\eqref{hdyn}, it is convenient to move to Fourier space, where tensor perturbations admit the decomposition~\cite{Watanabe:2006qe}
\begin{equation}
h_{ij}(t, \vec{x}) = \sum_{\lambda} \int \frac{d^3k}{(2\pi)^3} \, h^\lambda(t, \vec{k}) \, \epsilon^\lambda_{ij}(\vec{k}) \, e^{i \vec{k} \cdot \vec{x}}\,,
\end{equation}
where $\epsilon^\lambda_{ij}$ are the spin-2 polarization tensors. These satisfy the normalization condition $\sum_{ij} \epsilon^\lambda_{ij} \epsilon^{\lambda' *}_{ij} = 2 \delta^{\lambda \lambda'}$, with $\lambda = +, \times$ denoting the two independent polarization states of gravitational waves.

The mode function $h^\lambda(t, \vec{k})$ can be factorized as
\begin{equation}
h^\lambda(t, \vec{k}) = h_{\mathrm{prim}}^\lambda(\vec{k}) \, X(t, k)\,,
\end{equation}
where $k = |\vec{k}|$, $X(t,k)$ is the transfer function encoding the time evolution of each mode, and $h_{\mathrm{prim}}^\lambda(\vec{k})$ corresponds to the primordial tensor amplitude.

Within this framework, the tensor power spectrum is given by~\cite{Bernal:2020ywq}
\begin{equation}
\mathcal{P}_T(k) = \frac{k^3}{\pi^2} \sum_{\lambda} \left| h^\lambda_{\mathrm{prim}}(\vec{k}) \right|^2 = \frac{2}{\pi^2} \hspace{0.3mm} G \hspace{0.3mm} H^2 \Big|_{k = aH}\,,
\end{equation}
which shows that the amplitude of the spectrum is directly controlled by the Hubble parameter evaluated at horizon crossing. As a consequence, any modification to the background expansion history relative to the standard $\Lambda$CDM scenario is expected to leave observable signatures in the PGW spectrum. In the present framework, we assume that the leading effects of the generalized entropic corrections are encoded in the modified cosmological background evolution, while the primordial tensor generation mechanism retains its standard inflationary form.

In the absence of source terms, Eq.~\eqref{hdyn} can be recast into a simpler form resembling the equation of a damped harmonic oscillator. Expressed in terms of conformal time $\tau$, defined via $d\tau = dt/a$, it becomes
\begin{equation}
\label{Xd}
X'' + 2 \frac{a'}{a} X' + k^2 X = 0\,,
\end{equation}
where primes indicate derivatives with respect to $\tau$.

The energy density of PGWs generated at first order in tensor perturbations, within the standard cosmological framework, is given by~\cite{Watanabe:2006qe}
\begin{equation}
\Omega_{\mathrm{GW}}(\tau, k) = \frac{\left[ X'(\tau, k) \right]^2}{12 a^2(\tau) H^2(\tau)} \, \mathcal{P}_T(k) \simeq \left[ \frac{a_{\mathrm{hc}}}{a(\tau)} \right]^4 \left[ \frac{H_{\mathrm{hc}}}{H(\tau)} \right]^2 \frac{\mathcal{P}_T(k)}{24}\,,
\label{Ttps}
\end{equation}
where the second expression follows from averaging over many oscillation cycles. Under this approximation, one has
\begin{equation}
X'(\tau,k) \simeq k \, X(\tau,k) \simeq \frac{k \hspace{0.3mm} a_{\mathrm{hc}}}{\sqrt{2}\, a(\tau)} \simeq \frac{a_{\mathrm{hc}}^2 \hspace{0.3mm} H_{\mathrm{hc}}}{\sqrt{2}\, a(\tau)}\,,
\end{equation}
with the relation $k = 2\pi f = a_{\mathrm{hc}} H_{\mathrm{hc}}$ holding at horizon crossing.

This approximation is evaluated at the horizon-crossing epoch, defined as the moment when a mode with comoving wavenumber $k$ re-enters the Hubble radius, i.e., when its physical wavelength becomes comparable to the Hubble scale. At this point, the evolution of the mode starts to be significantly influenced by the background expansion, and its amplitude begins to decay due to Hubble damping. Therefore, horizon crossing provides a natural reference time for describing the subsequent evolution of gravitational wave modes.

Introducing the reduced Hubble parameter $h$, the present-day relic abundance of PGWs can be written as
\begin{equation}
\Omega_{\mathrm{GW}}(\tau_0, k) \, h^2 \simeq \left[ \frac{g_*(T_{\mathrm{hc}})}{2} \right] \left[ \frac{g_{*s}(T_0)}{g_{*s}(T_{\mathrm{hc}})} \right]^{4/3} \frac{\mathcal{P}_T(k) \, \Omega_r(T_0) \, h^2}{24}\,,
\label{Spt}
\end{equation}
where the radiation and entropy densities are expressed in terms of the effective relativistic degrees of freedom as $\rho(T) = \frac{\pi^2}{30} g_*(T) T^4$ and $s(T) = \frac{2\pi^2}{45} g_{*s}(T) T^3$, respectively. Here, $g_*(T)$ and $g_{*s}(T)$ quantify the number of relativistic species contributing to the energy and entropy content of the Universe at a given temperature.

In addition, the cosmological evolution is conveniently parametrized in terms of the temperature, which is related to the redshift via the standard relation $1+z = \dfrac{T}{T_0}$, where $T_0 \simeq 3\,\mathrm{K}$ denotes the current mean temperature of the Universe.

The scale dependence of the tensor power spectrum can be parametrized as~\cite{Watanabe:2006qe,Bernal:2020ywq}
\begin{equation}
\mathcal{P}_T(k) = A_T \left( \frac{k}{\tilde{k}} \right)^{n_T}\,,
\end{equation}
where $n_T$ is the tensor spectral index, and $\tilde{k} = 0.05\,\mathrm{Mpc}^{-1}$ is a commonly adopted pivot scale in cosmological analyses. The normalization $A_T$ is related to the amplitude of scalar perturbations $A_S$ through the tensor-to-scalar ratio $r$, such that $A_T = r A_S$.

\begin{figure}[t]
\centering
\includegraphics[width=16 cm]{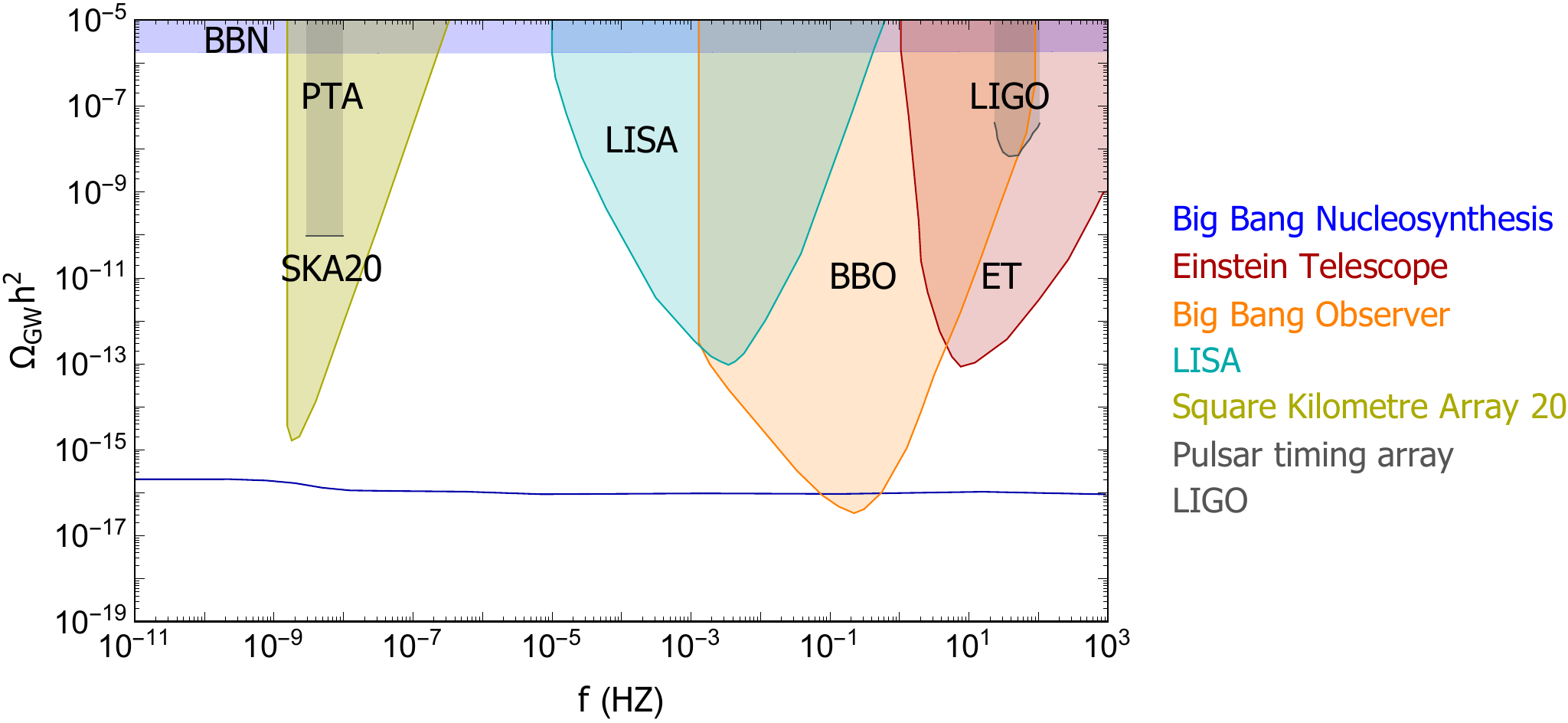}
\caption{Plot of the PGW spectrum as a function of the frequency $f$ for $n_T=0$, $A_S\simeq2.1\times10^{-9}$. The shaded regions represent the forecasted sensitivity ranges of various GW detectors~\cite{Breitbach:2018ddu}, which are listed on the right together with their expected launch years. In gray the regions excluded by PTA~\cite{KAGRA:2021kbb} and LIGO~\cite{Shannon:2015ect}.}
\label{Fig1}
\end{figure}

The relic abundance given in Eq.~\eqref{Spt} is shown as a function of the frequency $f$ in Fig.~\ref{Fig1} (blue solid curve), assuming a nearly scale-invariant primordial tensor spectrum ($n_T \simeq 0$) and normalized using the scalar perturbation amplitude inferred from Planck observations. In particular, the scalar amplitude at the pivot scale $\tilde{k}$ is constrained to $\ln(10^{10} A_S) = 3.044 \pm 0.014$, which corresponds to $A_S \simeq 2.1 \times 10^{-9}$~\cite{Planck:2018vyg}.

The shaded regions represent the projected sensitivity bands of several current and future gravitational wave detectors~\cite{Breitbach:2018ddu}, such as the LISA mission~\cite{amaro2017laser}, the Einstein Telescope (ET)~\cite{Sathyaprakash:2012jk}, the proposed Big Bang Observer (BBO)~\cite{Crowder:2005nr}, and the Square Kilometre Array (SKA)~\cite{Janssen:2014dka}. In addition, constraints arising from Big Bang Nucleosynthesis (BBN) are included, based on observational bounds on the effective number of relativistic neutrino species~\cite{Boyle:2007zx,Stewart:2007fu}. The gray shaded areas indicate regions of parameter space that are already ruled out by current measurements from Pulsar Timing Arrays (PTA)~\cite{KAGRA:2021kbb} and LIGO~\cite{Shannon:2015ect}.

\subsection{PGWs within the Kaniadakis entropy-based cosmology}

We now turn to the study of how the modified cosmological evolution described by Eq.~\eqref{ModHubRate} affects the PGW spectrum. Following Ref. \cite{Bernal:2020ywq}, throughout this analysis we assume that the underlying gravitational modifications mainly alter the background expansion history of the Universe, so that their effects can be effectively encoded in the modified Hubble rate. This approximation is well justified within the regime explored in the present work, where departures from standard general relativity remain perturbatively small.

To account for Kaniadakis-induced effects, it is useful to recast Eq.~\eqref{Ttps} into the form
\begin{eqnarray}
\nonumber
\Omega_{\mathrm{GW}}(\tau,k)
&=&
\left[\frac{a_{\mathrm{hc}}}{a(\tau)}\right]^4
\left[\frac{H_{\mathrm{hc}}}{H_{\mathrm{GR}}(\tau)}\right]^2
\left[\frac{H_{\mathrm{GR}}(\tau)}{H(\tau)}\right]^2
\frac{\mathcal{P}_T(k)}{24}
\\[2mm]
&=&
\Omega^{\mathrm{GR}}_{\mathrm{GW}}(\tau,k)
\left[\frac{H_{\mathrm{GR}}(\tau)}{H(\tau)}\right]^2
\left[\frac{a_{\mathrm{hc}}}{a_{\mathrm{hc}}^{\mathrm{GR}}}\right]^4
\left[\frac{a^{\mathrm{GR}}(\tau)}{a(\tau)}\right]^4
\left[\frac{H_{\mathrm{hc}}}{H_{\mathrm{hc}}^{\mathrm{GR}}}\right]^2 \,.
\label{eq:PGWBar0}
\end{eqnarray}
In this expression, quantities labeled by the superscript/subscript ``GR'' correspond to the standard cosmological evolution predicted within general relativity. In particular, $\Omega^{\mathrm{GR}}_{\mathrm{GW}}(\tau,k)$ represents the gravitational wave relic abundance obtained in the conventional framework, namely the result given in Eq.~\eqref{Ttps}. This decomposition is especially convenient because it isolates the corrections induced by the modified cosmological dynamics with respect to the standard GR prediction.

\begin{figure}[t]
\begin{center}
\includegraphics[width=16 cm]{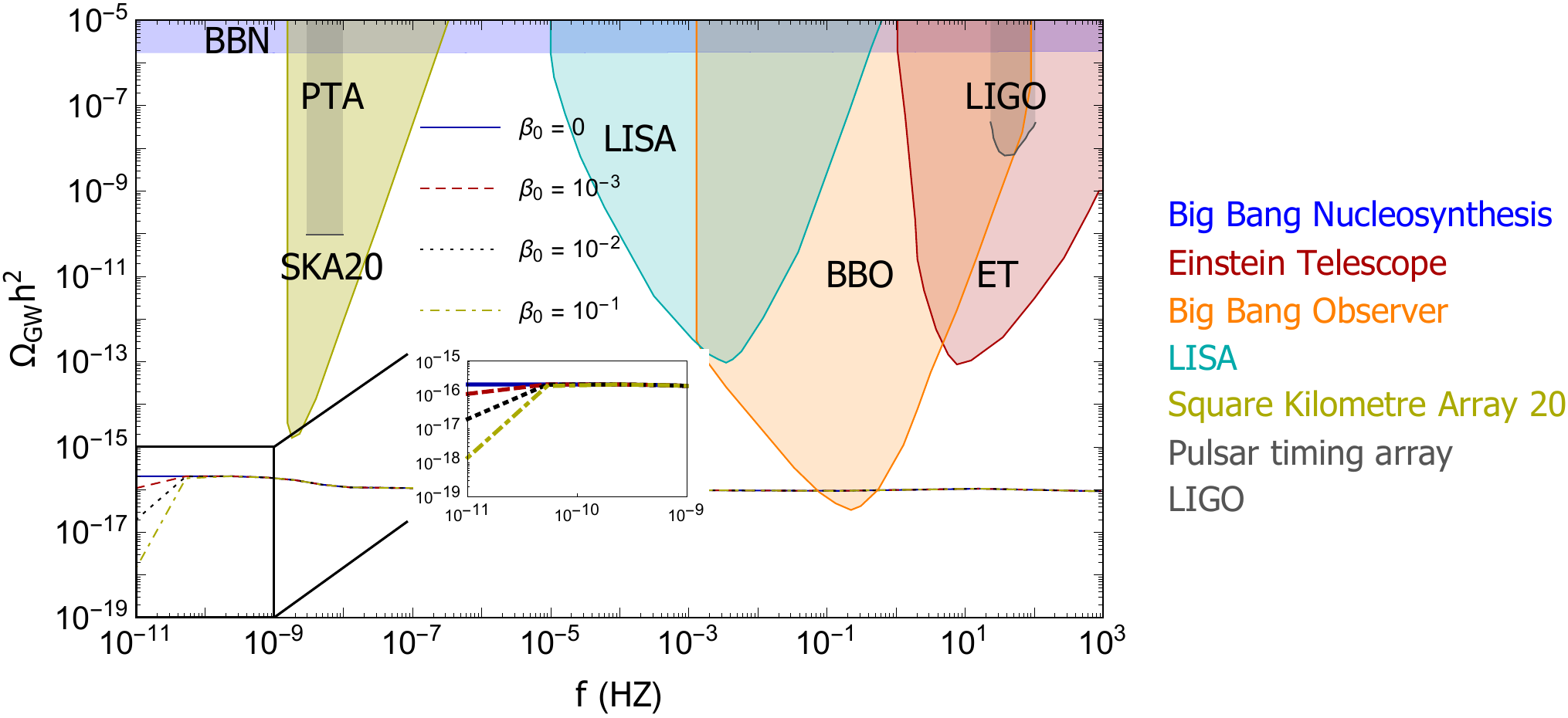}
\caption{Plot of the PGW spectrum as a function of the frequency $f$ for $n_T=0$, $A_S\simeq2.1\times10^{-9}$ and  different values of the entropic parameter $\beta_0$.}
\label{Fig2}
\end{center}
\end{figure}

By exploiting the condition that, at the present time, the modified cosmological scenario under consideration coincides with standard general relativity, we can finally rewrite Eq. \eqref{eq:PGWBar0} as
\begin{equation}
 \Omega_{\mathrm{GW}}(\tau_0,k)
= \Omega^{\mathrm{GR}}_{\mathrm{GW}}(\tau_0,k)\left[ \frac{a_{\mathrm{hc}}}{a_{\mathrm{hc}}^{\mathrm{GR}}}\right]^4\left[ \frac{H_{\mathrm{hc}}}{H_{\mathrm{hc}}^{\mathrm{GR}}}\right]^2   \,. \label{eq:PGWBar} 
\end{equation}
Figure~\ref{Fig2} displays the PGW spectrum as a function of the frequency \(f\), evaluated for different values of the rescaled entropic parameter
$\beta_0 \equiv \frac{\kappa \pi}{G H_0^2}$ (see Footnote~1). Overall, the deviations from the standard GR prediction remain extremely small across the entire frequency range considered. This behavior directly reflects the perturbative framework adopted in the present analysis, which requires $\beta_0 \ll 1$ and therefore motivates the choice of very small values of the Kaniadakis parameter, namely $\kappa \ll 10^{-123}$. 

Such a stringent regime does not arise from inflationary physics alone, but rather from the fact that the observable PGW spectrum depends not only on the primordial generation of tensor modes, but also on their subsequent propagation across the full post-inflationary expansion history of the Universe. Since the transfer function explicitly depends on the background evolution, assuming the modified Friedmann dynamics to remain valid up to the present epoch requires the perturbative expansion to hold at all cosmological times. Because the quantity $\beta(H)=\kappa\pi/(G H^2)$ reaches its maximum value today, where $H=H_0$, the resulting analysis is naturally restricted to tiny departures from the standard cosmological background.

Despite their small amplitude, the entropic corrections exhibit a characteristic frequency dependence, becoming more noticeable in the low-frequency regime and progressively suppressed toward higher frequencies.  This behavior can be directly understood from Eq.~\eqref{ModHubRate}:  since the correction term scales inversely with the Hubble parameter, its relative contribution rapidly decreases at early times, when the expansion rate of the Universe was much larger. As a consequence, high-frequency modes, which re-enter the horizon deep in the early Universe, evolve almost identically to the GR case, whereas low-frequency modes preserve a comparatively larger sensitivity to the modified background dynamics. 

More specifically, we find that the Kaniadakis corrections lead to a suppression of the PGW spectrum with respect to the standard GR prediction. While the resulting deviations 
are expected to remain below the projected sensitivity of forthcoming GW observatories, they may still provide physically interesting indications regarding possible cosmological applications of Kaniadakis statistical mechanics. In particular, the tiny values of $\kappa$ considered in the present analysis are fully consistent with the observational study of Ref.~\cite{Hernandez-Almada:2021aiw}, where Kaniadakis horizon-entropy cosmology was constrained using several independent cosmological datasets, including Cosmic Chronometers, Supernova Type Ia, HII galaxies, Strong Lensing Systems and Baryon Acoustic Oscillations observations.

From this perspective, the present PGW analysis may be regarded as a complementary probe of Kaniadakis-inspired cosmology, extending the phenomenological investigation of generalized entropy effects from late-time cosmological observables to the primordial tensor sector and the pre-BBN Universe. In this sense, PGWs offer a potentially valuable theoretical framework for testing whether generalized entropy-based modifications of gravity can consistently influence both the late- and early-time cosmological evolution while remaining compatible with the standard thermal history of the Universe and current observational constraints.

\subsection{Comparison with Other Extended Gravity Models}

It is worth emphasizing that modifications of the PGW spectrum also arise in several alternative cosmological and gravitational scenarios, making it important to distinguish the signatures predicted within the present framework from those associated with other extensions of standard cosmology. Since sizable anomalies in the primordial tensor spectrum are difficult to accommodate within the conventional FRW picture, they generally require suitable modifications either in the matter sector or in the underlying gravitational dynamics~\cite{Mavromatos}.

For instance, the presence of massive axion fields can generate characteristic triangular features in the PGW spectrum~\cite{Co:2021lkc}, whose structure encodes information about the axion potential and its dynamics during the early Universe. Similar signatures may also emerge in inflationary scenarios involving axion kination~\cite{Gouttenoire:2021wzu}. More generally, the implications of modified gravity theories for PGWs have been extensively investigated in different contexts; see Ref.~\cite{Odintsov:2022cbm} for a recent review.

In particular, scalar--tensor and extra-dimensional cosmologies were studied in Ref.~\cite{Bernal:2020ywq} by parameterizing the modified expansion history as $H(T)=A(T)\,H_{\rm GR}(T)$, where the amplification factor is commonly written as $A(T)=1+\eta\left(\frac{T}{T_*}\right)^\nu$.
Here, \(T_*\) denotes the characteristic temperature scale at which the corrections become relevant, while \(\eta\) and \(\nu\) are model-dependent parameters. Interestingly, the parameter \(\nu\) determines qualitatively different behaviors of the PGW spectrum. For \(\nu>0\), the resulting signal becomes blue-tilted relative to the standard GR prediction, leading to an enhancement at high frequencies. This occurs, for example, in Randall--Sundrum type II brane cosmology (\(\nu=2\)) and in kination scenarios (\(\nu=1\)). In the limiting case \(\nu=0\), the modification reduces to a constant amplification of the Hubble rate, producing an approximately frequency-independent enhancement of the PGW spectrum. Similar effects may also arise in the presence of a large number of extra relativistic degrees of freedom in the primordial plasma~\cite{PAMELA}. Conversely, scenarios with \(\nu<0\) typically generate localized features or bump-like structures around the reheating scale.

In comparison, as already highlighted above, the entropy-induced corrections associated with the generalized Kaniadakis framework become increasingly relevant in the low-frequency sector, while remaining strongly suppressed at high frequencies. This characteristic frequency dependence may therefore represent a potentially distinctive signature of Kaniadakis-inspired cosmology, differentiating it from scenarios typically characterized by blue-tilted spectra, localized bumps or resonant features.

Additional discriminating signatures may also arise beyond the spectral amplitude itself. In particular, extended gravity theories and generalized cosmological frameworks can modify the polarization content of gravitational waves, potentially inducing scalar or vector polarization modes, polarization mixing, birefringence effects or modified tensor amplitudes. Along this line, recent studies have shown that future space-based interferometers such as LISA could achieve remarkable sensitivity to deviations from the standard GR polarization structure, providing stringent constraints on non-tensorial modes and modified propagation effects~\cite{Akama:2026}. In this sense, combining the analysis of the PGW spectral shape with future polarization measurements may provide a promising strategy to disentangle entropy-induced cosmological effects from other classes of modified gravity scenarios.

\section{Slow-roll inflation in Kaniadakis Cosmology}
\label{infl}

In this section, we examine inflation within the Kaniadakis-modified Friedmann cosmology. Following \cite{keskin2022inflationary}, we focus on the high-energy phase of the early Universe under the slow-roll approximation \cite{odintsov2017inflationary}, assuming that the dynamics is driven by a scalar field $\phi$ with potential $V(\phi)$.

Contrary to the PGW analysis discussed previously, the present investigation is restricted to the inflationary epoch only. Consequently, the validity of the perturbative expansion in Eq.~\eqref{modified_friedmann_flat} must be imposed at the inflationary energy scale, namely for $H \sim H_{\rm inf}$, rather than over the entire cosmological evolution. Since the perturbative parameter scales as $\beta(H)=\kappa\pi/(G H^2)$, the consistency of the perturbative treatment requires $\beta(H_{\rm inf}) \ll 1$ for a typical inflationary scale $H_{\rm inf}\sim10^{-5}M_{\rm Pl}$, where $M_{\rm Pl}$ is the reduced Planck mass. This condition implies $\kappa < 1.3 \times 10^{-12}\sim \mathcal{O}(10^{-12})$, thus allowing the Kaniadakis parameter to attain comparatively larger values while still preserving the perturbative consistency of the theory. As a consequence, the generalized entropic corrections are expected to play a more relevant role during the inflationary era, potentially leading to deviations from the standard inflationary predictions that could be phenomenologically more significant and observationally testable.

The key inflationary observables 
are the tensor-to-scalar ratio $r$, the scalar spectral index $n_s$ and the running of the scalar spectral index which, 
for minimal coupling, are given by \cite{nojiri2017modified, hwang2005classical}:
	\begin{align}
    \label{nsind}
		r &= 16 \, \epsilon, \\[2mm]
		n_s &= 1 - 6 \, \epsilon + 2 \, \eta, \\[2mm]
		\alpha_{s}&\equiv \frac{d n_{s}}{d\ln k}
\simeq -\,\frac{d n_{s}}{dN}, 
	\end{align}
where \(N\) is the number of e-folds before the end of inflation. Here \(k\) denotes the comoving wavenumber of the perturbation mode, related to the comoving wavelength by \(k=2\pi/\lambda\). We used the horizon-crossing relation \(k=aH\), which during slow-roll inflation implies $d\ln k \simeq -\, dN$. Furthermore, the inflationary parameters defined as
	\begin{equation}
		\epsilon = - \frac{\dot H}{H^2}\,, \qquad
		\eta = - \frac{\ddot H}{2 H \dot H}\,.
	\end{equation}
Assuming a canonical scalar-field description of the inflationary sector, the corresponding Lagrangian density can be written as
$\mathcal{L}=X-V(\phi)$, where $X=-\frac{1}{2}g^{\mu\nu}\partial_{\mu}\phi\,\partial_{\nu}\phi$ denotes the kinetic term, while \(V(\phi)\) represents the scalar-field potential. Under the assumption of spatial homogeneity, the associated energy density and pressure of the inflaton field are given by
$\rho_{\phi}=\frac{\dot{\phi}^{2}}{2}+V(\phi)$, $p_{\phi}=\frac{\dot{\phi}^{2}}{2}-V(\phi)$. 
The corresponding scalar-field fluid satisfies the continuity equation, which can equivalently be rewritten as the Klein-Gordon equation $\ddot{\phi}+3H\dot{\phi}+\partial_\phi V=0$. 

Assuming that the potential energy dominates, the slow-roll conditions take the form \cite{nojiri2017modified,alhallak2023salvaging},
	\begin{equation}
		\ddot{\phi} \ll H \dot{\phi}, \qquad \frac{\dot{\phi}^2}{2} \ll V(\phi).
	\end{equation}
This leads to the following dynamics \cite{Lambiase:2023ryq},
	\begin{equation}
		\dot{\phi}\simeq-\dfrac{1}{3H}\partial_{\phi} V. \label{envolv}
	\end{equation}
Solving the modified Friedmann equation (\ref{modified2}) (with $\Lambda=0$) for $H$ and applying the above approximation, one finds to leading order in $\kappa$ \cite{Lambiase:2023ryq},
	\begin{equation}
		H \simeq \sqrt{\frac{8\pi G}{3} V} + \sqrt{\frac{27 \pi}{2  G^{7}V^{3}}}\,\dfrac{\kappa^2}{64} , \label{hubble}
	\end{equation}
	where the explicit dependence $V(\phi)$ is suppressed for brevity. 
    
    In Ref. \cite{Lambiase:2023ryq}, it has been shown that the evolution of Hubble parameter and inflationary parameters for  Kaniadakis entropy can be written as
\begin{align}
	\dot{H} &\simeq 
	\left(-4\pi G + \frac{9 \pi \kappa^2}{32 G^3 V^2} \right) \dot{\phi}^2 , \label{kiu} \\[2mm]
	\epsilon &\simeq 
	\left( \frac{3}{2V} - \frac{27 \kappa^2}{128} \frac{1}{G^4 V^3} \right) \dot{\phi}^2 , \label{kiuss} \\[2mm]
	\eta &\simeq 
	\left[
	-\sqrt{ \frac{3}{8 \pi G V}} \ddot{\phi} 
	+ \sqrt{\frac{3}{2 \pi G^ 9 V^7}} \frac{9 \kappa^2}{512} 
	\left( \ddot{\phi} V - 2 \dot{\phi}^2 \frac{\partial V}{\partial \phi} \right)
	\right] \frac{1}{\dot{\phi}}\, . \label{eta}
\end{align}

The extent of inflation is quantified by the number of e-folds $N$, defined as \cite{remmen2014many},
\begin{equation}
	N = \int_{t_i}^{t_f} H(t)\, dt ,
\end{equation}
where $t_i$ and $t_f$ represent the initial and final times of the inflationary phase, respectively.

Since primordial inflaton fluctuations become relevant at horizon crossing, we identify the initial time with the horizon-crossing time, namely $t_i\equiv t_c$, corresponding to $\phi_i\equiv\phi_c$. During slow-roll evolution one has $\dot{\phi}<0$ and hence $\phi_c>\phi_f$. Therefore, the number of e-folds can be equivalently written as
\begin{equation}
	N = -\int_{\phi_c}^{\phi_f} \frac{H}{\dot{\phi}}\, d\phi
	= \int_{\phi_f}^{\phi_c} \frac{H}{|\dot{\phi}|}\, d\phi .
\end{equation}

Contrary to Ref.~\cite{Lambiase:2023ryq}, where inflation was investigated by adopting power-law scalar-field potentials, in the present work we focus on a Starobinsky-like inflationary scenario~\cite{Starobinsky1980}. Similar analyses based on power-law inflationary potentials have also been recently considered in the framework of (dual) Kaniadakis cosmology~\cite{Liravi:2026xzm}. Our choice of the Starobinsky potential is motivated both by its remarkable phenomenological success and by its natural connection with modified gravity and quantum-corrected cosmological frameworks. In particular, the model predicts values of the scalar spectral index and tensor-to-scalar ratio that are in excellent agreement with the latest Planck observations, while naturally supporting a sufficiently long slow-roll phase without requiring significant fine-tuning of the parameters.

Accordingly, we consider the potential~\cite{piattella2018lecture}
\begin{equation}
V(\phi)=\dfrac{3M^{2} M_{\text{Pl}}^{2}}{4}\left[1-\exp\left(-\sqrt{\dfrac{2}{3}}\frac{\phi}{M_{\text{Pl}}}\right)\right]^{2},
\end{equation}
where $M_{\text{Pl}}$ fixes the gravitational scale of the theory, while $M$ determines the characteristic energy scale of inflation. The Starobinsky model can be interpreted as an effective description arising from quadratic corrections to the Einstein--Hilbert action, which naturally emerge when quantum fields are coupled to a classical gravitational background~\cite{piattella2018lecture}. In the following, we focus on the large-field regime $\phi \gg M_{\mathrm{Pl}}$, where the slow-roll conditions are satisfied.

For the present model, and by using the results (\ref{envolv}) and (\ref{hubble}), one can easily show that the field evolves with time as follows
\begin{equation}
\dot{\phi}\simeq 
-\frac{M}{2\sqrt{3\pi G}}
\exp\left(-\sqrt{\frac{2}{3}}\frac{\phi}{M_{\rm Pl}}\right)
+
\frac{2\pi^{3/2}\kappa^2}{G^{5/2}M^3\sqrt{3}}\,
\frac{
\exp\left(-\sqrt{\frac{2}{3}}\frac{\phi}{M_{\rm Pl}}\right)
}{
\left[
1-\exp\left(-\sqrt{\frac{2}{3}}\frac{\phi}{M_{\rm Pl}}\right)
\right]^4
}\, .
\end{equation}
Therefore, one can show that
\begin{equation}
\epsilon \simeq 
\frac{4}{3}
\frac{
\exp\left(-2\sqrt{\frac{2}{3}}\frac{\phi}{M_{\rm Pl}}\right)
}{
\left[
1-\exp\left(-\sqrt{\frac{2}{3}}\frac{\phi}{M_{\rm Pl}}\right)
\right]^2
}
-
\frac{32\kappa^2\pi^2}{G^2M^4}
\frac{
\exp\left(-2\sqrt{\frac{2}{3}}\frac{\phi}{M_{\rm Pl}}\right)
}{
\left[
1-\exp\left(-\sqrt{\frac{2}{3}}\frac{\phi}{M_{\rm Pl}}\right)
\right]^6
}\, .
\end{equation}
The Kaniadakis parameter $\kappa$ introduces a negative correction to the slow-roll parameter $\epsilon$, thus modifying the exit dynamics with respect to the standard Starobinsky scenario.

Let us impose that inflation (and the counting of e-folds $N$) terminates when the slow-roll parameter satisfies $\epsilon(\phi_f) \sim 1$ \cite{Guth:1980zm, Starobinsky1980,Linde:1983gd}. One gets
\begin{equation}
\phi_f =
\sqrt{\frac{3}{2}}\,M_{\rm Pl}
\left[
-\ln(2\sqrt{3}-3)
-\left(39+\frac{45\sqrt{3}}{2}\right)
\frac{\kappa^2\pi^2}{G^2M^4}
\right] .
\end{equation}
For $\phi \gg M_{\rm Pl}$, the number of e-folds is
\begin{align}
N \simeq\;&
\frac{3}{4}
\left[
\exp\left(\sqrt{\frac{2}{3}}\frac{\phi}{M_{\rm Pl}}\right)\right]
+
\frac{6\pi^{2}\kappa^{2}}{G^{2}M^{4}}
\Bigg\{
\exp\left(\sqrt{\frac{2}{3}}\frac{\phi}{M_{\rm Pl}}\right)
+3\ln\left[
\exp\left(\sqrt{\frac{2}{3}}\frac{\phi}{M_{\rm Pl}}\right)-1
\right]
\nonumber\\[2mm]
&
-\frac{3}{
\exp\left(\sqrt{\frac{2}{3}}\frac{\phi}{M_{\rm Pl}}\right)-1
}
-\frac{1}{
2\left[
\exp\left(\sqrt{\frac{2}{3}}\frac{\phi}{M_{\rm Pl}}\right)-1
\right]^2
}
\Bigg\} ,
\label{Nefold}
\end{align}
where we have retained only the leading Starobinsky contribution in the large-\(N\) regime, corresponding to the dominant exponential term, so that the standard result is consistently recovered in the limit \(\kappa \to 0\).
Here, $\phi$ denotes the scalar-field value at horizon crossing, while the contribution evaluated at the end of inflation has been neglected since it is subleading in the large-field regime $\phi \gg M_{\rm Pl}$.

For the slow-roll parameter $\eta$ in Eq.~\eqref{eta}, one obtains
\begin{equation}
\eta \simeq
-\frac{4}{3}
\frac{
\exp\left(-\sqrt{\frac{2}{3}}\frac{\phi}{M_{\rm Pl}}\right)
}{
1-\exp\left(-\sqrt{\frac{2}{3}}\frac{\phi}{M_{\rm Pl}}\right)
}
+
\frac{32\kappa^2\pi^2}{3G^2M^4}
\frac{
\exp\left(-2\sqrt{\frac{2}{3}}\frac{\phi}{M_{\rm Pl}}\right)
\left[
3+\exp\left(\sqrt{\frac{2}{3}}\frac{\phi}{M_{\rm Pl}}\right)
\right]
}{
\left[
1-\exp\left(-\sqrt{\frac{2}{3}}\frac{\phi}{M_{\rm Pl}}\right)
\right]^6
}\, .
\end{equation}
Therefore, the scalar spectral index can be expressed as
\begin{equation}
n_s \simeq
1-
\frac{
8\exp\left(-2\sqrt{\frac{2}{3}}\frac{\phi}{M_{\rm Pl}}\right)
\left[
2+\exp\left(\sqrt{\frac{2}{3}}\frac{\phi}{M_{\rm Pl}}\right)
\right]
}{
3\left[
1-\exp\left(-\sqrt{\frac{2}{3}}\frac{\phi}{M_{\rm Pl}}\right)
\right]^2
}
+
\frac{
64\kappa^2\pi^2
\exp\left(-2\sqrt{\frac{2}{3}}\frac{\phi}{M_{\rm Pl}}\right)
\left[
12+\exp\left(\sqrt{\frac{2}{3}}\frac{\phi}{M_{\rm Pl}}\right)
\right]
}{
3G^2M^4
\left[
1-\exp\left(-\sqrt{\frac{2}{3}}\frac{\phi}{M_{\rm Pl}}\right)
\right]^6
}\, .
\end{equation}
For large $N$, one obtains (see Appendix \ref{AppA})
\begin{equation}
\label{nsap}
n_s \simeq
1-\frac{2}{N}
+
\frac{\kappa^2\pi^2}{G^2M^4N^2}
\left[
120-36\ln\left(\frac{4N}{3}\right)
\right] ,
\end{equation}
where we have retained the leading Starobinsky contribution in the large-\(N\) expansion, together with the leading correction induced by the Kaniadakis parameter at order \(\kappa^2\).

Considering the phenomenologically relevant value \(N\simeq 60\), typically associated with horizon crossing in standard slow-roll inflationary scenarios~\cite{piattella2018lecture}, Eq.~\eqref{nsap} shows that the scalar spectral index receives a negative correction induced by the Kaniadakis parameter $\kappa$. In particular, the contribution proportional to $\kappa^2$ shifts $n_s$ toward smaller values, corresponding to a slightly larger deviation from exact scale invariance and therefore to an enhancement of the red tilt of the primordial power spectrum. Such a behavior originates from the generalized entropy corrections encoded in the modified cosmological dynamics, which affect both the Hubble evolution and the slow-roll parameters.

\begin{figure}[t]
\centering

\begin{minipage}{0.49\textwidth}
    \centering
    \includegraphics[width=\linewidth]{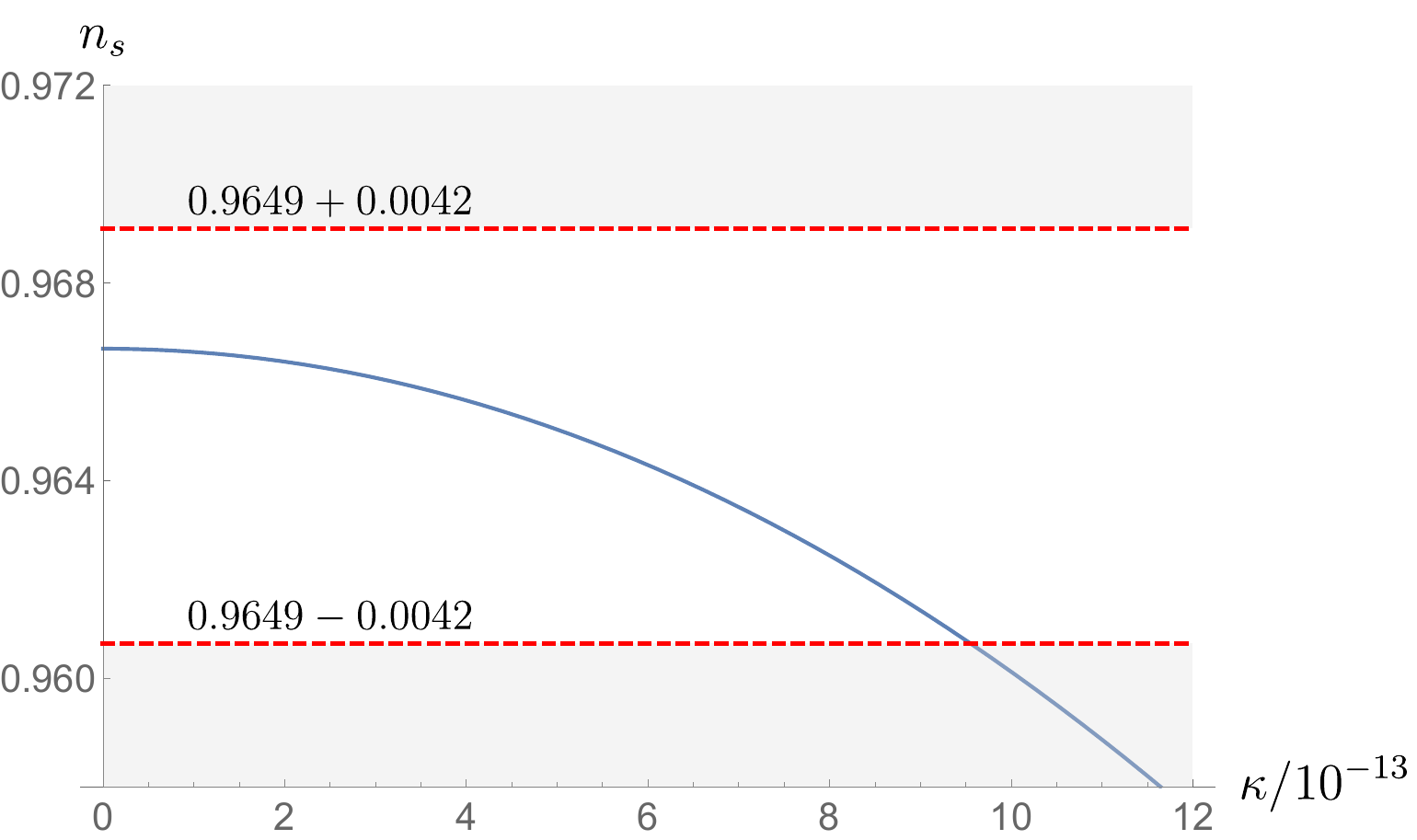}
\end{minipage}
\hfill
\begin{minipage}{0.49\textwidth}
    \centering
    \includegraphics[width=\linewidth]{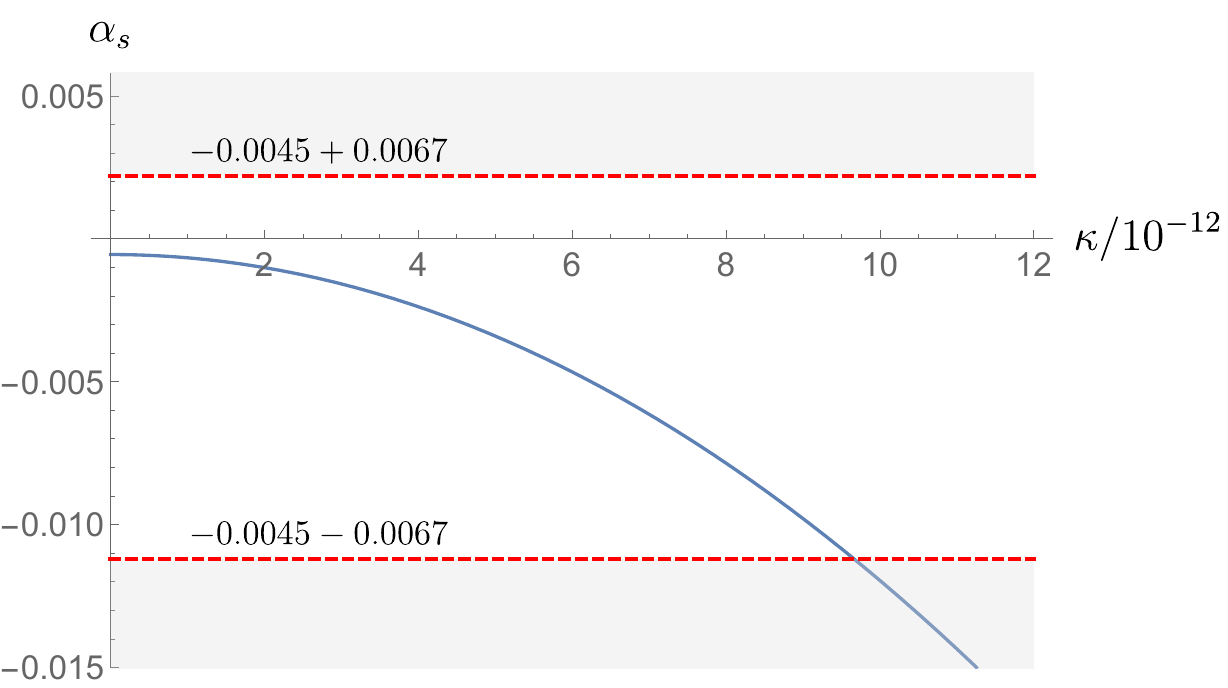}
\end{minipage}

\vspace{0.5cm}

\begin{minipage}{0.49\textwidth}
    \centering
    \includegraphics[width=\linewidth]{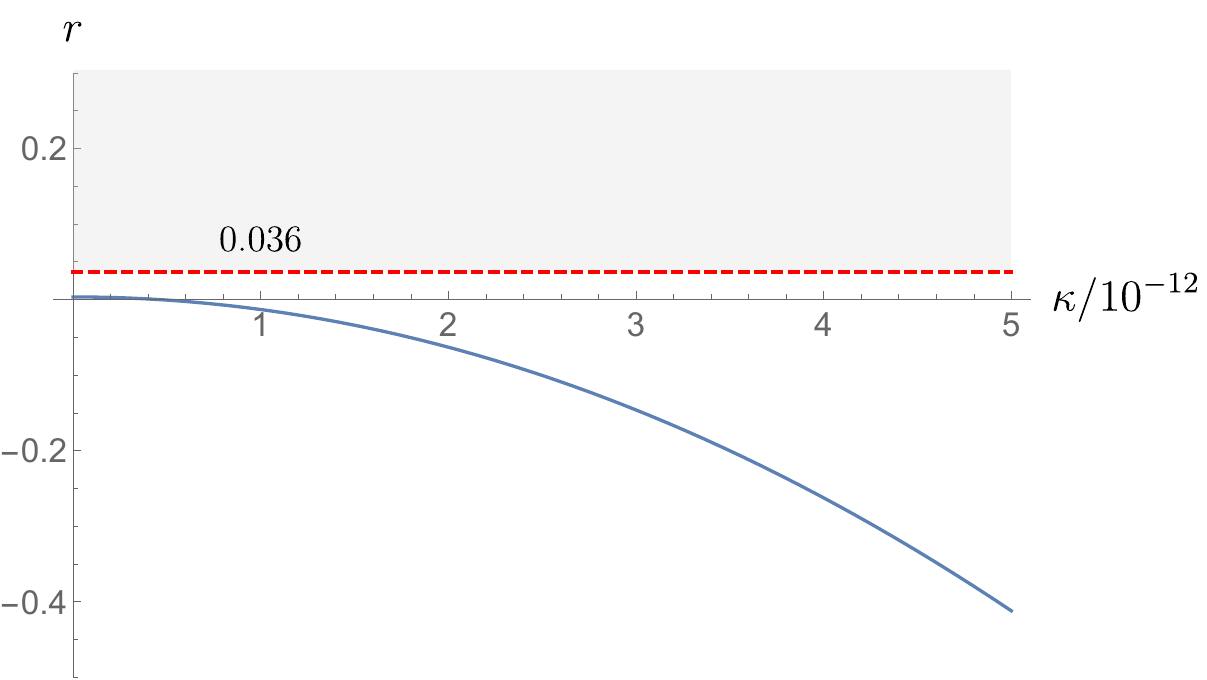}
\end{minipage}

\caption{Evolution of the inflationary observables as functions of the rescaled Kaniadakis parameter. Upper-left panel: scalar spectral index $n_s$. Upper-right panel: running of the scalar spectral index $\alpha_s$. Lower panel: tensor-to-scalar ratio $r$. The blue (red) lines denote the corresponding theoretical predictions (observational bounds), while the shaded gray regions indicate the excluded parameter space. 
}
\label{nsplot}
\end{figure}

Nevertheless, the resulting deviations from the standard Starobinsky prediction remain strongly constrained by the current observational bounds on the scalar spectral index. Indeed, for such parameter, the latest Planck observations provide the constraint~\cite{Planck:2018vyg}
\begin{equation}
n_s = 0.9649 \pm 0.0042
\hspace{0.5cm}
(68\%\,\, \mathrm{C.L.}) .
\end{equation}
Assuming an inflaton mass of the order \(M \sim 10^{-5} M_{\mathrm{Pl}}\), with its value fixed by the observed amplitude of the CMB fluctuations~\cite{ivanov2022analytic}, one obtains the following bound on the Kaniadakis index:
\begin{equation}
\kappa \lesssim 9.6 \times 10^{-13}\,, \label{ns}
\end{equation}
where we have restricted the analysis to positive values of \(\kappa\), following the discussion below Eq.~\eqref{KenBH}. We notice that the upper observational limit on \(n_s\) is automatically satisfied throughout the physically allowed parameter region, whereas the lower Planck bound yields the nontrivial constraint reported above (see the upper-left panel of Fig. \ref{nsplot}). Interestingly, the resulting bound on $\kappa$ is slightly more restrictive than the perturbative consistency requirement discussed at the beginning of this Section, although it remains of the same order of magnitude. This indicates that current CMB observations are already probing the boundary of the perturbatively controlled parameter space of the model.

On the other hand, by substituting the large-$N$ expression of the scalar spectral index, we find
\begin{equation}
\alpha_s \simeq
-\frac{2}{N^2}
+
\frac{\kappa^2\pi^2}{G^2M^4N^3}
\left[
276-72\ln\left(\frac{4N}{3}\right)
\right] .
\label{spectral}
\end{equation}
This result shows that the Kaniadakis correction introduces a negative contribution to the running of the scalar spectral index. As a consequence, the magnitude of the standard negative Starobinsky prediction is slightly enhanced, shifting $\alpha_s$ toward more negative values and therefore strengthening the scale-dependent distortion of the primordial scalar spectrum.

From the latest Planck observations, the running of the spectral index is constrained to be \cite{akrami2020planck}
\begin{equation}
\alpha_s = -0.0045 \pm 0.0067 \hspace{0.5cm} (68\%\,\, C.L.) \,.
\end{equation}
For $N\simeq60$, and restricting to positive values of $\kappa$ as discussed below Eq.~\eqref{KenBH}, the lower observational bound on $\alpha_s$ provides the nontrivial constraint (see the upper-right panel of Fig. \ref{nsplot})
\begin{equation}
\kappa \lesssim 9.7\times10^{-12}\,.
\label{alpha}
\end{equation}
The upper observational limit is instead automatically satisfied in the allowed parameter range. It is worth emphasizing that the above constraint is weaker than the perturbative consistency requirement discussed previously and therefore does not further restrict the physically viable parameter space. Nevertheless, it shows that the entire perturbatively consistent regime of the model remains fully compatible with current observational bounds on the running of the scalar spectral index.

Finally, for the tensor-to-scalar ratio, one obtains
\begin{equation}
r \simeq 
\frac{64}{3}
\frac{
\exp\left(-2\sqrt{\frac{2}{3}}\frac{\phi}{M_{\rm Pl}}\right)
}{
\left[
1-\exp\left(-\sqrt{\frac{2}{3}}\frac{\phi}{M_{\rm Pl}}\right)
\right]^2
}
-
\frac{512\kappa^{2}\pi^{2}}{G^{2}M^{4}}
\frac{
\exp\left(-2\sqrt{\frac{2}{3}}\frac{\phi}{M_{\rm Pl}}\right)
}{
\left[
1-\exp\left(-\sqrt{\frac{2}{3}}\frac{\phi}{M_{\rm Pl}}\right)
\right]^6
}\, .
\end{equation}
In terms of $N$, and by using the approximation $N\gg1$, one gets
\begin{equation}
r \simeq 
\frac{12}{N^2}
-
\frac{96\pi^2\kappa^2}{G^2M^4N^2}\,.
\end{equation}
From the latest BICEP/Keck and Planck observations, the tensor-to-scalar ratio is constrained as~\cite{BICEP:2021xfz}
\begin{equation}
r < 0.036 \hspace{0.5cm} (95\%\,\, \mathrm{C.L.}) .
\end{equation}
For \(N\simeq 60\), the standard Starobinsky prediction already gives
\(r\simeq 3.3\times 10^{-3}\), well below the observational upper limit. Since the Kaniadakis correction further decreases the tensor-to-scalar ratio, suppressing the relative amplitude of primordial tensor perturbations with respect to scalar perturbations, \(r\) does not provide an additional nontrivial constraint on the Kaniadakis parameter in the present setup (see the lower panel of Fig.~\ref{nsplot}).

Therefore, among the inflationary observables analyzed in the present framework, the scalar spectral index provides the strongest and phenomenologically relevant constraint on the Kaniadakis parameter, yielding an upper bound close to the limit imposed by perturbative consistency. The corresponding constraint obtained from the running of the scalar spectral index remains of the same order of magnitude, although comparatively less restrictive and fully compatible with the perturbatively allowed parameter region.


Remarkably, the bound obtained within the Starobinsky-like inflationary scenario considered here is consistent with the one derived in Ref.~\cite{Lambiase:2023ryq} in the context of slow-roll inflation driven by power-law scalar-field potentials. Despite the different structure of the inflationary potentials and the distinct background dynamics characterizing the two models, the resulting constraints on the Kaniadakis parameter remain of comparable magnitude. This suggests that the observational bounds on \(\kappa\) are relatively robust with respect to the specific realization of the inflationary sector, and mainly originate from the characteristic features of the generalized entropy corrections on the slow-roll evolution and on the primordial perturbation spectra. 

On the other hand, substantially different estimates for the Kaniadakis parameter have been obtained from late-time cosmological analyses~\cite{Hernandez-Almada:2021rjs, Luciano:2025ykr,Blasone:2023yke,Luciano:2022knb,Sheykhi:2025zre,Salehi:2024kkj}. However, such bounds are not directly comparable to the present inflationary result, since they probe very different physical regimes and energy scales of the cosmological evolution.

From a broader perspective, this apparent scale dependence may actually suggest a more general physical picture in which the Kaniadakis parameter effectively evolves with the cosmological epoch, namely $\kappa\equiv\kappa(z)$. Such a possibility is well motivated within the holographic interpretation of relativistic Kaniadakis statistics, where entropy encodes the effective microscopic degrees of freedom of the Universe. In this framework, one may naturally expect relativistic entropic effects to be more relevant at high energies and in the earliest stages of the cosmic evolution, while progressively approaching the standard Boltzmann-Gibbs limit at late times. In this sense, the inflationary bound obtained here may be interpreted as probing the high-energy behavior of a potentially dynamical Kaniadakis coupling, whereas late-time cosmological observations constrain its low-energy limit. This scenario could provide a unified framework connecting generalized entropy effects across different cosmological epochs.


\section{Conclusions and Outlook}
\label{Conc}

In this work, we have investigated the cosmological implications of Kaniadakis entropy within two complementary and observationally relevant frameworks, namely PGWs and slow-roll inflation in a Starobinsky-like scenario. Starting from the generalized entropy-area relation associated with Kaniadakis statistics, we derived the corresponding modified Friedmann equations and analyzed how the resulting entropic corrections affect both the background expansion history of the Universe and the evolution of primordial cosmological observables.

First, we studied the spectrum of PGWs in the Kaniadakis cosmological framework. By incorporating the modified background dynamics induced by the generalized entropy corrections, we evaluated the resulting relic gravitational-wave spectrum and compared it with the standard cosmological prediction. Our analysis shows that the Kaniadakis corrections induce characteristic deviations in the low-frequency sector, while remaining progressively suppressed at higher frequencies. Although these effects remain perturbatively small within the regime considered here, they provide a distinctive and theoretically well-motivated signature of entropy-based cosmological modifications. 

Subsequently, we analyzed slow-roll inflation by considering a Starobinsky-inspired inflationary potential within the Kaniadakis-modified Friedmann framework. We derived the corrected slow-roll dynamics and obtained explicit analytical expressions for the scalar spectral index, its running and the tensor-to-scalar ratio. Our results show that generalized entropy corrections can induce nontrivial modifications to the inflationary observables, thereby establishing a direct connection between horizon thermodynamics and the physics of the primordial Universe. By confronting the theoretical predictions with the latest Planck and BICEP/Keck observations, we derived stringent bounds on the Kaniadakis parameter. In particular, the scalar spectral index provides the strongest phenomenological constraint, while the running yields a less restrictive but still complementary bound. The tensor-to-scalar ratio remains well below the current observational upper limit throughout the perturbatively allowed parameter region.

An important outcome of the present analysis is that the obtained constraints are broadly consistent with previous investigations of Kaniadakis cosmology performed in different inflationary scenarios. This suggests that the resulting bounds are not merely artifacts of a specific model realization, but rather reflect a model-independent tendency of generalized entropy corrections on cosmological dynamics. In particular, the fact that comparable constraints emerge from distinct inflationary frameworks supports the idea that the Kaniadakis parameter effectively controls general modifications of the slow-roll evolution and of the primordial perturbation spectra, independently of the detailed structure of the inflationary potential.

Several directions deserve further investigation. First, extending the present analysis beyond the slow-roll approximation could clarify how Kaniadakis corrections affect the evolution of scalar perturbations, especially near the end of inflation where such effects become more significant. 
Another interesting perspective concerns primordial non-Gaussianities. Since generalized entropy corrections modify the inflationary dynamics, they may also induce observable deviations from the nearly Gaussian statistics predicted in standard inflationary models, providing additional phenomenological signatures beyond $n_s$, $r$, and $\alpha_s$.
Finally, it would be worthwhile to investigate stochastic gravitational-wave backgrounds generated during cosmological phase transitions, particularly the QCD phase transition. In this context, the modified expansion history induced by Kaniadakis entropy could leave potentially detectable imprints on the resulting gravitational-wave spectrum. Studies along these directions are currently underway and will be presented in future work.

\appendix

\section{Perturbative derivation of the scalar spectral index}
\label{AppA}
In this appendix, we present the perturbative calculation of the scalar spectral index $n_s$ in our model. To this end, starting from Eq. \eqref{Nefold}, let us define 
\begin{equation} 
Y=\exp\left(\sqrt{\frac{2}{3}}\frac{\phi}{M_{\rm Pl}}\right), \qquad K=\frac{\kappa^2\pi^2}{G^2M^4}. 
\end{equation}
The e-fold relation we are using takes the form 
\begin{equation} 
N \simeq \frac{3}{4}Y +6K\left[ Y+3\ln(Y-1)-\frac{3}{Y-1} -\frac{1}{2(Y-1)^2} \right]. 
\end{equation}
We invert it perturbatively by using 
\begin{equation}
\label{Yex}
Y=Y_0+K\,Y_1. 
\end{equation}
At zeroth order, we then have
\begin{equation}
N=\frac{3}{4}Y_0 \quad\Rightarrow\quad Y_0=\frac{4N}{3},
\end{equation} 
while, at first order in \(K\), 
it follows that 
\begin{align}
\nonumber
&0=\frac{3}{4}Y_1 +6\left[ Y_0+3\ln(Y_0-1)-\frac{3}{Y_0-1} -\frac{1}{2(Y_0-1)^2} \right] \\[2mm]
&\Longrightarrow\,\, Y_1= -8\left[ Y_0+3\ln(Y_0-1)-\frac{3}{Y_0-1} -\frac{1}{2(Y_0-1)^2} \right]. 
\end{align}

Now, starting from the definition \eqref{nsind}
and using
\begin{eqnarray}
\epsilon&=&
\frac{4}{3}\frac{Y^{-2}}{(1-Y^{-1})^2}
-
K\frac{32\,Y^{-2}}{(1-Y^{-1})^6}\,,\\[2mm]
\eta&=&
-\frac{4}{3}\frac{Y^{-1}}{1-Y^{-1}}
+
K\,\frac{32}{3}
\frac{Y^{-2}(3+Y)}{(1-Y^{-1})^6}\,,
\end{eqnarray}
one obtains
\begin{equation}
n_s(Y)=
1-\frac{8(Y+2)}{3(Y-1)^2}
+
K\,\frac{64Y^4(Y+12)}{3(Y-1)^6}. \label{A8}
\end{equation}
Substituting Eq.~(\ref{Yex}) into Eq.~(\ref{A8}), we perform a perturbative expansion around the zeroth-order solution $Y_0$, retaining terms up to first order in $K$:
\begin{equation}
n_s(Y_0 + K Y_1)
\simeq
n_s(Y_0)
+
K Y_1
\left.
\frac{d n_s}{dY}
\right|_{Y_0}.
\end{equation}
Taking subsequently the large-$N$ limit, we obtain
\begin{equation}
n_s \simeq
1-\frac{2}{N}
+
\frac{K}{N^2}
\left[
120-36\ln\left(\frac{4N}{3}\right)
\right].
\end{equation}
Restoring \(K=\kappa^2\pi^2/(G^2M^4)\), this becomes
\begin{equation}
n_s \simeq
1-\frac{2}{N}
+
\frac{\kappa^2\pi^2}{G^2M^4N^2}
\left[
120-36\ln\left(\frac{4N}{3}\right)
\right].
\end{equation}
The usual Starobinsky result is consistently recovered in the limit \(\kappa\to0\).

\acknowledgments 
The research of GGL is supported by the postdoctoral fellowship program of the University of Lleida. GGL gratefully acknowledges the contribution of the LISA Cosmology Working Group (CosWG), as well as support from the COST Actions CA21136 - \textit{Addressing observational tensions in cosmology with systematics and fundamental physics (CosmoVerse)} - CA23130, \textit{Bridging high and low energies in search of quantum gravity (BridgeQG)} and CA21106 -  \textit{COSMIC WISPers in the Dark Universe: Theory, astrophysics and experiments (CosmicWISPers)}.

\bibliographystyle{apsrev4-1}
\bibliography{kan}

@article{Bekenstein:1973ur,
    author = "Bekenstein, Jacob D.",
    title = "{Black holes and entropy}",
    doi = "10.1103/PhysRevD.7.2333",
    journal = "Phys. Rev. D",
    volume = "7",
    pages = "2333--2346",
    year = "1973"
}

@article{Barman:2023ktz,
    author = "Barman, Basabendu and Ghoshal, Anish and Grzadkowski, Bohdan and Socha, Anna",
    title = "{Measuring inflaton couplings via primordial gravitational waves}",
    eprint = "2305.00027",
    archivePrefix = "arXiv",
    primaryClass = "hep-ph",
    doi = "10.1007/JHEP07(2023)231",
    journal = "JHEP",
    volume = "07",
    pages = "231",
    year = "2023"
}

@article{amaro2017laser,
  title={Laser interferometer space antenna},
  author={Amaro-Seoane, Pau and others},
  journal={arXiv:1702.00786},
  year={2017}
}

@article{Sathyaprakash:2012jk,
    author = "Sathyaprakash, B. and others",
    editor = "Hannam, Mark and Sutton, Patrick and Hild, Stefan and van den Broeck, Chris",
    title = "{Scientific Objectives of Einstein Telescope}",
    eprint = "1206.0331",
    archivePrefix = "arXiv",
    primaryClass = "gr-qc",
    doi = "10.1088/0264-9381/29/12/124013",
    journal = "Class. Quant. Grav.",
    volume = "29",
    pages = "124013",
    year = "2012",
    note = "[Erratum: Class.Quant.Grav. 30, 079501 (2013)]"
}

@article{Stewart:2007fu,
    author = "Stewart, Andrew and Brandenberger, Robert",
    title = "{Observational Constraints on Theories with a Blue Spectrum of Tensor Modes}",
    eprint = "0711.4602",
    archivePrefix = "arXiv",
    primaryClass = "astro-ph",
    doi = "10.1088/1475-7516/2008/08/012",
    journal = "JCAP",
    volume = "08",
    pages = "012",
    year = "2008"
}

@article{Boyle:2007zx,
    author = "Boyle, Latham A. and Buonanno, Alessandra",
    title = "{Relating gravitational wave constraints from primordial nucleosynthesis, pulsar timing, laser interferometers, and the CMB: Implications for the early Universe}",
    eprint = "0708.2279",
    archivePrefix = "arXiv",
    primaryClass = "astro-ph",
    doi = "10.1103/PhysRevD.78.043531",
    journal = "Phys. Rev. D",
    volume = "78",
    pages = "043531",
    year = "2008"
}

@article{Janssen:2014dka,
    author = "Janssen, Gemma and others",
    editor = "Bourke, Tyler L. and others",
    title = "{Gravitational wave astronomy with the SKA}",
    eprint = "1501.00127",
    archivePrefix = "arXiv",
    primaryClass = "astro-ph.IM",
    doi = "10.22323/1.215.0037",
    journal = "PoS",
    volume = "AASKA14",
    pages = "037",
    year = "2015"
}

@article{Shannon:2015ect,
    author = "Shannon, R. M. and others",
    title = "{Gravitational waves from binary supermassive black holes missing in pulsar observations}",
    eprint = "1509.07320",
    archivePrefix = "arXiv",
    primaryClass = "astro-ph.CO",
    doi = "10.1126/science.aab1910",
    journal = "Science",
    volume = "349",
    number = "6255",
    pages = "1522--1525",
    year = "2015"
}

@article{KAGRA:2021kbb,
    author = "Abbott, R. and others",
    collaboration = "KAGRA, Virgo, LIGO Scientific",
    title = "{Upper limits on the isotropic gravitational-wave background from Advanced LIGO and Advanced Virgo{\textquoteright}s third observing run}",
    eprint = "2101.12130",
    archivePrefix = "arXiv",
    primaryClass = "gr-qc",
    reportNumber = "LIGO-DCC-P2000314",
    doi = "10.1103/PhysRevD.104.022004",
    journal = "Phys. Rev. D",
    volume = "104",
    number = "2",
    pages = "022004",
    year = "2021"
}

@article{Crowder:2005nr,
    author = "Crowder, Jeff and Cornish, Neil J.",
    title = "{Beyond LISA: Exploring future gravitational wave missions}",
    eprint = "gr-qc/0506015",
    archivePrefix = "arXiv",
    doi = "10.1103/PhysRevD.72.083005",
    journal = "Phys. Rev. D",
    volume = "72",
    pages = "083005",
    year = "2005"
}

@article{Breitbach:2018ddu,
    author = "Breitbach, Moritz and Kopp, Joachim and Madge, Eric and Opferkuch, Toby and Schwaller, Pedro",
    title = "{Dark, Cold, and Noisy: Constraining Secluded Hidden Sectors with Gravitational Waves}",
    eprint = "1811.11175",
    archivePrefix = "arXiv",
    primaryClass = "hep-ph",
    reportNumber = "CERN-TH-2018-255, MITP/18-115",
    doi = "10.1088/1475-7516/2019/07/007",
    journal = "JCAP",
    volume = "07",
    pages = "007",
    year = "2019"
}

@article{Maity:2024cpq,
    author = "Maity, Suvashis and Haque, Md Riajul",
    title = "{Probing the early universe with future GW observatories}",
    eprint = "2407.18246",
    archivePrefix = "arXiv",
    primaryClass = "astro-ph.CO",
    doi = "10.1088/1475-7516/2025/04/091",
    journal = "JCAP",
    volume = "04",
    pages = "091",
    year = "2025"
}

@article{Bernal:2020ywq,
    author = "Bernal, Nicol{\'a}s and Ghoshal, Anish and Hajkarim, Fazlollah and Lambiase, Gaetano",
    title = "{Primordial Gravitational Wave Signals in Modified Cosmologies}",
    eprint = "2008.04959",
    archivePrefix = "arXiv",
    primaryClass = "gr-qc",
    doi = "10.1088/1475-7516/2020/11/051",
    journal = "JCAP",
    volume = "11",
    pages = "051",
    year = "2020"
}

@article{Watanabe:2006qe,
    author = "Watanabe, Yuki and Komatsu, Eiichiro",
    title = "{Improved Calculation of the Primordial Gravitational Wave Spectrum in the Standard Model}",
    eprint = "astro-ph/0604176",
    archivePrefix = "arXiv",
    doi = "10.1103/PhysRevD.73.123515",
    journal = "Phys. Rev. D",
    volume = "73",
    pages = "123515",
    year = "2006"
}

@article{Efstathiou:2020wem,
    author = "Efstathiou, George and Gratton, Steven",
    title = "{The evidence for a spatially flat Universe}",
    eprint = "2002.06892",
    archivePrefix = "arXiv",
    primaryClass = "astro-ph.CO",
    doi = "10.1093/mnrasl/slaa093",
    journal = "Mon. Not. Roy. Astron. Soc.",
    volume = "496",
    number = "1",
    pages = "L91--L95",
    year = "2020"
}

@article{Bekenstein:1974ax,
    author = "Bekenstein, Jacob D.",
    title = "{Generalized second law of thermodynamics in black hole physics}",
    doi = "10.1103/PhysRevD.9.3292",
    journal = "Phys. Rev. D",
    volume = "9",
    pages = "3292--3300",
    year = "1974"
}

@article{Hawking:1975vcx,
    author = "Hawking, S. W.",
    editor = "Gibbons, G. W. and Hawking, S. W.",
    title = "{Particle Creation by Black Holes}",
    doi = "10.1007/BF02345020",
    journal = "Commun. Math. Phys.",
    volume = "43",
    pages = "199--220",
    year = "1975",
    note = "[Erratum: Commun.Math.Phys. 46, 206 (1976)]"
}

@article{Padmanabhan:2003gd,
    author = "Padmanabhan, T.",
    title = "{Gravity and the thermodynamics of horizons}",
    eprint = "gr-qc/0311036",
    archivePrefix = "arXiv",
    doi = "10.1016/j.physrep.2004.10.003",
    journal = "Phys. Rept.",
    volume = "406",
    pages = "49--125",
    year = "2005"
}

@article{Cai:2005ra,
    author = "Cai, Rong-Gen and Kim, Sang Pyo",
    title = "{First law of thermodynamics and Friedmann equations of Friedmann-Robertson-Walker universe}",
    eprint = "hep-th/0501055",
    archivePrefix = "arXiv",
    doi = "10.1088/1126-6708/2005/02/050",
    journal = "JHEP",
    volume = "02",
    pages = "050",
    year = "2005"
}

@article{Srednicki:1993im,
    author = "Srednicki, Mark",
    title = "{Entropy and area}",
    eprint = "hep-th/9303048",
    archivePrefix = "arXiv",
    reportNumber = "LBL-33754, CFPA-93-02",
    doi = "10.1103/PhysRevLett.71.666",
    journal = "Phys. Rev. Lett.",
    volume = "71",
    pages = "666--669",
    year = "1993"
}

@article{NC,
  author  = {Wilk, G. and Wlodarczyk, Z.},
  title   = {Interpretation of the Nonextensivity Parameter q in Some Applications of Tsallis Statistics and L\'evy Distributions},
  journal = {Physical Review D},
  volume  = {50},
  number  = {4},
  pages   = {2318--2322},
  year    = {1994},
  doi     = {10.1103/PhysRevD.50.2318}
}

@article{plasma,
  author  = {Hasegawa, A. and Mima, K. and Duong-van, M.},
  title   = {Stationary Spectrum of Strong Turbulence in Magnetized Nonuniform Plasma},
  journal = {Physical Review Letters},
  volume  = {54},
  pages   = {2608--2610},
  year    = {1985},
  doi     = {10.1103/PhysRevLett.54.2608}
}

@article{OSC,
  author  = {Carvalho, J. C. and Silva, R. and do Nascimento Jr., J. D. and Soares, B. B. and De Medeiros, J. R.},
  title   = {Nonextensive effects on oscillatory systems},
  journal = {Europhysics Letters (EPL)},
  volume  = {91},
  number  = {6},
  pages   = {69002},
  year    = {2010},
  doi     = {10.1209/0295-5075/91/69002}
}

@article{Jeans2,
  author  = {Abreu, E. M. C. and Neto, J. A. and Barboza, E. M. and Nunes, R. C.},
  title   = {Tsallis and Kaniadakis statistics from the viewpoint of entropic gravity formalism},
  journal = {International Journal of Modern Physics A},
  volume  = {32},
  number  = {17},
  pages   = {1750028},
  year    = {2017},
  doi     = {10.1142/S0217751X17500281}
}

@article{Jeans1,
  author  = {Abreu, E. M. C. and Ananias Neto, J. and Barboza, E. M. and Nunes, R. C.},
  title   = {Jeans instability criterion from the viewpoint of Kaniadakis' statistics},
  journal = {Europhysics Letters (EPL)},
  volume  = {114},
  pages   = {55001},
  year    = {2016},
  doi     = {10.1209/0295-5075/114/55001}
}

@article{Kaniadakis:2002zz,
    author = "Kaniadakis, G.",
    title = "{Statistical mechanics in the context of special relativity}",
    eprint = "cond-mat/0210467",
    archivePrefix = "arXiv",
    doi = "10.1103/PhysRevE.66.056125",
    journal = "Phys. Rev. E",
    volume = "66",
    pages = "056125",
    year = "2002"
}

@article{Carlip:2000nv,
    author = "Carlip, Steven",
    title = "{Logarithmic corrections to black hole entropy from the Cardy formula}",
    eprint = "gr-qc/0005017",
    archivePrefix = "arXiv",
    reportNumber = "UCD-00-11, UCD-2000-11",
    doi = "10.1088/0264-9381/17/20/302",
    journal = "Class. Quant. Grav.",
    volume = "17",
    pages = "4175--4186",
    year = "2000"
}

@article{Abbott:1984fp,
    author = "Abbott, L. F. and Wise, Mark B.",
    title = "{Constraints on Generalized Inflationary Cosmologies}",
    reportNumber = "CALT-68-1100",
    doi = "10.1016/0550-3213(84)90329-8",
    journal = "Nucl. Phys. B",
    volume = "244",
    pages = "541--548",
    year = "1984"
}

@article{Albrecht:1982wi,
    author = "Albrecht, Andreas and Steinhardt, Paul J.",
    editor = "Fang, Li-Zhi and Ruffini, R.",
    title = "{Cosmology for Grand Unified Theories with Radiatively Induced Symmetry Breaking}",
    reportNumber = "UPR-0185T",
    doi = "10.1103/PhysRevLett.48.1220",
    journal = "Phys. Rev. Lett.",
    volume = "48",
    pages = "1220--1223",
    year = "1982"
}

@article{Rubakov:1982df,
    author = "Rubakov, V. A. and Sazhin, M. V. and Veryaskin, A. V.",
    title = "{Graviton Creation in the Inflationary Universe and the Grand Unification Scale}",
    doi = "10.1016/0370-2693(82)90641-4",
    journal = "Phys. Lett. B",
    volume = "115",
    pages = "189--192",
    year = "1982"
}

@article{Linde:1983gd,
    author = "Linde, Andrei D.",
    title = "{Chaotic Inflation}",
    doi = "10.1016/0370-2693(83)90837-7",
    journal = "Phys. Lett. B",
    volume = "129",
    pages = "177--181",
    year = "1983"
}

@article{Grishchuk:1974ny,
    author = "Grishchuk, L. P.",
    title = "{Amplification of gravitational waves in an isotropic universe}",
    journal = "Sov. Phys. JETP",
    volume = "40",
    number = "3",
    pages = "409--415",
    year = "1975"
}

@article{Starobinsky:1979ty,
    author = "Starobinsky, Alexei A.",
    editor = "Khalatnikov, I. M. and Mineev, V. P.",
    title = "{Spectrum of relict gravitational radiation and the early state of the universe}",
    journal = "JETP Lett.",
    volume = "30",
    pages = "682--685",
    year = "1979"
}

@article{Guth:1980zm,
    author = "Guth, Alan H.",
    editor = "Fang, Li-Zhi and Ruffini, R.",
    title = "{The Inflationary Universe: A Possible Solution to the Horizon and Flatness Problems}",
    reportNumber = "SLAC-PUB-2576",
    doi = "10.1103/PhysRevD.23.347",
    journal = "Phys. Rev. D",
    volume = "23",
    pages = "347--356",
    year = "1981"
}

@article{Tsallis:2013,
    author = "Tsallis, Constantino and Cirto, Leonardo J. L.",
    title = "{Black hole thermodynamical entropy}",
    doi = "10.1140/epjc/s10052-013-2487-6",
    journal = "Eur. Phys. J. C",
    volume = "73",
    number = "7",
    year = "2013"
}

@article{Barrow:2020tzx,
    author = "Barrow, John D.",
    title = "{The Area of a Rough Black Hole}",
    eprint = "2004.09444",
    archivePrefix = "arXiv",
    primaryClass = "gr-qc",
    doi = "10.1016/j.physletb.2020.135643",
    journal = "Phys. Lett. B",
    volume = "808",
    pages = "135643",
    year = "2020"
}

@article{Dagotto:1989gp,
    author = "Dagotto, Elbio and Kocic, Aleksandar and Kogut, John B.",
    title = "{COLLAPSE OF THE WAVE FUNCTION, ANOMALOUS DIMENSIONS AND CONTINUUM LIMITS IN MODEL SCALAR FIELD THEORIES}",
    reportNumber = "ILL-TH-89-58, NSF-ITP-90-11",
    doi = "10.1016/0370-2693(90)91442-E",
    journal = "Phys. Lett. B",
    volume = "237",
    pages = "268--273",
    year = "1990"
}

@inproceedings{renyi1961entropy,
  title={On measures of entropy and information},
  author={Renyi, Alfred},
  booktitle={Proceedings of the Fourth Berkeley Symposium on Mathematical Statistics and Probability},
  volume={1},
  pages={547--561},
  year={1961},
  publisher={University of California Press}
}

@article{Kaniadakis2001non,
    author = "Kaniadakis, G.",
    title = "{Non-linear kinetics underlying generalized statistics}",
    eprint = "cond-mat/0103467",
    archivePrefix = "arXiv",
    doi = "10.1016/s0378-4371(01)00184-4",
    journal = "Physica A",
    volume = "296",
    number = "3-4",
    pages = "405--425",
    year = "2001"
}

@article{Jizba:2024klq,
    author = "Jizba, Petr and Lambiase, Gaetano and Luciano, Giuseppe Gaetano and Mastrototaro, Leonardo",
    title = "{Imprints of Barrow\textendash{}Tsallis cosmology in primordial gravitational waves}",
    eprint = "2403.09797",
    archivePrefix = "arXiv",
    primaryClass = "gr-qc",
    doi = "10.1140/epjc/s10052-024-13455-5",
    journal = "Eur. Phys. J. C",
    volume = "84",
    number = "10",
    pages = "1076",
    year = "2024"
}

@article{Jacobson:1995ab,
    author = "Jacobson, Ted",
    title = "{Thermodynamics of space-time: The Einstein equation of state}",
    eprint = "gr-qc/9504004",
    archivePrefix = "arXiv",
    reportNumber = "UMDGR-95-114",
    doi = "10.1103/PhysRevLett.75.1260",
    journal = "Phys. Rev. Lett.",
    volume = "75",
    pages = "1260--1263",
    year = "1995"
}

@article{Padmanabhan:2009vy,
    author = "Padmanabhan, T.",
    title = "{Thermodynamical Aspects of Gravity: New insights}",
    eprint = "0911.5004",
    archivePrefix = "arXiv",
    primaryClass = "gr-qc",
    doi = "10.1088/0034-4885/73/4/046901",
    journal = "Rept. Prog. Phys.",
    volume = "73",
    pages = "046901",
    year = "2010"
}

@article{Frolov:2002va,
    author = "Frolov, Andrei V. and Kofman, Lev",
    title = "{Inflation and de Sitter thermodynamics}",
    eprint = "hep-th/0212327",
    archivePrefix = "arXiv",
    reportNumber = "CITA-2002-46",
    doi = "10.1088/1475-7516/2003/05/009",
    journal = "JCAP",
    volume = "05",
    pages = "009",
    year = "2003"
}

@article{Akbar:2006kj,
    author = "Akbar, M. and Cai, Rong-Gen",
    title = "{Thermodynamic Behavior of Friedmann Equations at Apparent Horizon of FRW Universe}",
    eprint = "hep-th/0609128",
    archivePrefix = "arXiv",
    doi = "10.1103/PhysRevD.75.084003",
    journal = "Phys. Rev. D",
    volume = "75",
    pages = "084003",
    year = "2007"
}

@article{Akbar:2006er,
    author = "Akbar, M. and Cai, Rong-Gen",
    title = "{Friedmann equations of FRW universe in scalar-tensor gravity, f(R) gravity and first law of thermodynamics}",
    eprint = "hep-th/0602156",
    archivePrefix = "arXiv",
    doi = "10.1016/j.physletb.2006.02.035",
    journal = "Phys. Lett. B",
    volume = "635",
    pages = "7--10",
    year = "2006"
}

@article{Luciano:2024mcn,
    author = "Luciano, Giuseppe Gaetano and Sheykhi, Ahmad",
    title = "{Cosmology from string T-duality and zero-point length}",
    eprint = "2404.12707",
    archivePrefix = "arXiv",
    primaryClass = "gr-qc",
    doi = "10.1140/epjc/s10052-024-13035-7",
    journal = "Eur. Phys. J. C",
    volume = "84",
    number = "7",
    pages = "682",
    year = "2024"
}

@article{Sheykhi:2025zre,
    author = "Sheykhi, Ahmad and Sooraki, Ava Shahbazi and Liravi, Leila",
    title = "{Big bang nucleosynthesis constraints on dual Kaniadakis cosmology}",
    eprint = "2504.21146",
    archivePrefix = "arXiv",
    primaryClass = "gr-qc",
    doi = "10.1103/fg96-fjnw",
    journal = "Phys. Rev. D",
    volume = "112",
    number = "10",
    pages = "103546",
    year = "2025"
}

@article{Drepanou:2021jiv,
    author = "Drepanou, Niki and Lymperis, Andreas and Saridakis, Emmanuel N. and Yesmakhanova, Kuralay",
    title = "{Kaniadakis holographic dark energy and cosmology}",
    eprint = "2109.09181",
    archivePrefix = "arXiv",
    primaryClass = "gr-qc",
    doi = "10.1140/epjc/s10052-022-10415-9",
    journal = "Eur. Phys. J. C",
    volume = "82",
    number = "5",
    pages = "449",
    year = "2022"
}

@article{Luciano:2023bai,
    author = "Luciano, Giuseppe Gaetano and Saridakis, Emmanuel",
    title = "{P {\ensuremath{-}} v criticalities, phase transitions and geometrothermodynamics of charged AdS black holes from Kaniadakis statistics}",
    eprint = "2308.12669",
    archivePrefix = "arXiv",
    primaryClass = "gr-qc",
    doi = "10.1007/JHEP12(2023)114",
    journal = "JHEP",
    volume = "12",
    pages = "114",
    year = "2023"
}

@article{Gibbons:1977mu,
    author = "Gibbons, G. W. and Hawking, S. W.",
    title = "{Cosmological Event Horizons, Thermodynamics, and Particle Creation}",
    doi = "10.1103/PhysRevD.15.2738",
    journal = "Phys. Rev. D",
    volume = "15",
    pages = "2738--2751",
    year = "1977"
}

@article{Cai:2009zp,
    author = "Cai, Yi-Fu and Saridakis, Emmanuel N. and Setare, Mohammad R. and Xia, Jun-Qing",
    title = "{Quintom Cosmology: Theoretical implications and observations}",
    eprint = "0909.2776",
    archivePrefix = "arXiv",
    primaryClass = "hep-th",
    doi = "10.1016/j.physrep.2010.04.001",
    journal = "Phys. Rept.",
    volume = "493",
    pages = "1--60",
    year = "2010"
}

@article{Bartolo:2004if,
    author = "Bartolo, N. and Komatsu, E. and Matarrese, Sabino and Riotto, A.",
    title = "{Non-Gaussianity from inflation: Theory and observations}",
    eprint = "astro-ph/0406398",
    archivePrefix = "arXiv",
    reportNumber = "DFPD-04-A-12",
    doi = "10.1016/j.physrep.2004.08.022",
    journal = "Phys. Rept.",
    volume = "402",
    pages = "103--266",
    year = "2004"
}

@article{Bombelli:1986rw,
    author = "Bombelli, Luca and Koul, Rabinder K. and Lee, Joohan and Sorkin, Rafael D.",
    title = "{A Quantum Source of Entropy for Black Holes}",
    reportNumber = "PRINT-86-0371 (SYRACUSE)",
    doi = "10.1103/PhysRevD.34.373",
    journal = "Phys. Rev. D",
    volume = "34",
    pages = "373--383",
    year = "1986"
}

@article{Kaul:2000kf,
    author = "Kaul, Romesh K. and Majumdar, Parthasarathi",
    title = "{Logarithmic correction to the Bekenstein-Hawking entropy}",
    eprint = "gr-qc/0002040",
    archivePrefix = "arXiv",
    doi = "10.1103/PhysRevLett.84.5255",
    journal = "Phys. Rev. Lett.",
    volume = "84",
    pages = "5255--5257",
    year = "2000"
}

@article{Olive:1989nu,
    author = "Olive, Keith A.",
    title = "{Inflation}",
    reportNumber = "UMN-TH-804-89",
    doi = "10.1016/0370-1573(90)90144-Q",
    journal = "Phys. Rept.",
    volume = "190",
    pages = "307--403",
    year = "1990"
}

@book{CANTATA:2021asi,
    author = "Akrami, Yashar and others",
    editor = "Saridakis, Emmanuel N. and Lazkoz, Ruth and Salzano, Vincenzo and Vargas Moniz, Paulo and Capozziello, Salvatore and Beltr{\'a}n Jim{\'e}nez, Jose and De Laurentis, Mariafelicia and Olmo, Gonzalo J.",
    collaboration = "CANTATA",
    title = "{Modified Gravity and Cosmology. An Update by the CANTATA Network}",
    eprint = "2105.12582",
    archivePrefix = "arXiv",
    primaryClass = "gr-qc",
    doi = "10.1007/978-3-030-83715-0",
    isbn = "978-3-030-83714-3, 978-3-030-83717-4, 978-3-030-83715-0",
    publisher = "Springer",
    year = "2021"
}

@article{Capozziello:2011et,
    author = "Capozziello, Salvatore and De Laurentis, Mariafelicia",
    title = "{Extended Theories of Gravity}",
    eprint = "1108.6266",
    archivePrefix = "arXiv",
    primaryClass = "gr-qc",
    doi = "10.1016/j.physrep.2011.09.003",
    journal = "Phys. Rept.",
    volume = "509",
    pages = "167--321",
    year = "2011"
}

@article{Planck:2018vyg,
    author = "Aghanim, N. and others",
    collaboration = "Planck",
    title = "{Planck 2018 results. VI. Cosmological parameters}",
    eprint = "1807.06209",
    archivePrefix = "arXiv",
    primaryClass = "astro-ph.CO",
    doi = "10.1051/0004-6361/201833910",
    journal = "Astron. Astrophys.",
    volume = "641",
    pages = "A6",
    year = "2020",
    note = "[Erratum: Astron.Astrophys. 652, C4 (2021)]"
}

@article{Lymperis:2021qty,
    author = "Lymperis, Andreas and Basilakos, Spyros and Saridakis, Emmanuel 
N.",
    title = "{Modified cosmology through Kaniadakis horizon entropy}",
    eprint = "2108.12366",
    archivePrefix = "arXiv",
    primaryClass = "gr-qc",
    doi = "10.1140/epjc/s10052-021-09852-9",
    journal = "Eur. Phys. J. C",
    volume = "81",
    number = "11",
    pages = "1037",
    year = "2021"
}

@article{Mavromatos,
    author = "Mavromatos, Nick E. and Spanos, Vassilis C. and Stamou, Ioanna D.",
    title = "{Primordial black holes and gravitational waves in multiaxion-Chern-Simons inflation}",
    eprint = "2206.07963",
    archivePrefix = "arXiv",
    primaryClass = "hep-th",
    reportNumber = "KCL-PH-TH/2022-36",
    doi = "10.1103/PhysRevD.106.063532",
    journal = "Phys. Rev. D",
    volume = "106",
    number = "6",
    pages = "063532",
    year = "2022"
}

@article{Akama:2026,
    author = "Akama, Shingo and others",
    collaboration = "LISA Cosmology Working Group",
    title = "{Testing gravitational wave polarizations with LISA}",
    eprint = "2603.03165",
    archivePrefix = "arXiv",
    primaryClass = "astro-ph.CO",
    month = "3",
    year = "2026"
}

@article{PAMELA,
    author = "Catena, R. and others",
    title = "{Thermal Relics in Modified Cosmologies: Bounds on Evolution Histories of the Early Universe and Cosmological Boosts for PAMELA}",
    eprint = "0912.4421",
    archivePrefix = "arXiv",
    primaryClass = "astro-ph.CO",
    reportNumber = "DFTT-71-2009",
    doi = "10.1103/PhysRevD.81.123522",
    journal = "Phys. Rev. D",
    volume = "81",
    pages = "123522",
    year = "2010"
}

@article{Odintsov:2022cbm,
    author = "Odintsov, Sergei D. and Oikonomou, Vasilis K. and Myrzakulov, Ratbay",
    title = "{Spectrum of Primordial Gravitational Waves in Modified Gravities: A Short Overview}",
    eprint = "2204.00876",
    archivePrefix = "arXiv",
    primaryClass = "gr-qc",
    doi = "10.3390/sym14040729",
    journal = "Symmetry",
    volume = "14",
    number = "4",
    pages = "729",
    year = "2022"
}

@article{Gouttenoire:2021wzu,
    author = "Gouttenoire, Yann and Servant, G{\'e}raldine and Simakachorn, Peera",
    title = "{Revealing the Primordial Irreducible Inflationary Gravitational-Wave Background with a Spinning Peccei-Quinn Axion}",
    eprint = "2108.10328",
    archivePrefix = "arXiv",
    primaryClass = "hep-ph",
    reportNumber = "DESY 21-126",
    month = "8",
    year = "2021"
}

@article{Co:2021lkc,
    author = "Co, Raymond T. and Dunsky, David and Fernandez, Nicolas and Ghalsasi, Akshay and Hall, Lawrence J. and Harigaya, Keisuke and Shelton, Jessie",
    title = "{Gravitational wave and CMB probes of axion kination}",
    eprint = "2108.09299",
    archivePrefix = "arXiv",
    primaryClass = "hep-ph",
    reportNumber = "UMN-TH-4023/21, FTPI-MINN-21-15, CERN-TH-2021-124",
    doi = "10.1007/JHEP09(2022)116",
    journal = "JHEP",
    volume = "09",
    pages = "116",
    year = "2022"
}

@article{Luciano:2022knb,
    author = "Luciano, Giuseppe Gaetano",
    title = "{Modified Friedmann equations from Kaniadakis entropy and cosmological implications on baryogenesis and ${}^7 Li$-abundance}",
    doi = "10.1140/epjc/s10052-022-10285-1",
    journal = "Eur. Phys. J. C",
    volume = "82",
    number = "4",
    pages = "314",
    year = "2022"
}

@article{Blasone:2023yke,
    author = "Blasone, Massimo and Lambiase, Gaetano and Luciano, Giuseppe Gaetano",
    title = "{Kaniadakis entropy-based characterization of IceCube PeV neutrino signals}",
    eprint = "2309.16732",
    archivePrefix = "arXiv",
    primaryClass = "physics.gen-ph",
    doi = "10.1016/j.dark.2023.101342",
    journal = "Phys. Dark Univ.",
    volume = "42",
    pages = "101342",
    year = "2023"
}

@article{Luciano:2025ykr,
    author = "Luciano, Giuseppe Gaetano and Paliathanasis, Andronikos",
    title = "{Late-time cosmological constraints on Kaniadakis holographic dark energy}",
    eprint = "2509.17527",
    archivePrefix = "arXiv",
    primaryClass = "astro-ph.CO",
    doi = "10.1140/epjc/s10052-025-15122-9",
    journal = "Eur. Phys. J. C",
    volume = "85",
    number = "12",
    pages = "1384",
    year = "2025"
}

@article{Hernandez-Almada:2021rjs,
    author = "Hern\'andez-Almada, A. and Leon, Genly and Maga\~na, Juan and 
Garc\'\i{}a-Aspeitia, Miguel A. and Motta, V. and Saridakis, Emmanuel N. and 
Yesmakhanova, Kuralay and Millano, Alfredo D.",
    title = "{Observational constraints and dynamical analysis of Kaniadakis 
horizon-entropy cosmology}",
    eprint = "2112.04615",
    archivePrefix = "arXiv",
    primaryClass = "astro-ph.CO",
    doi = "10.1093/mnras/stac795",
    journal = "Mon. Not. Roy. Astron. Soc.",
    volume = "512",
    number = "4",
    pages = "5122--5134",
    year = "2022"
}

@article{Lambiase:2023ryq,
    author = "Lambiase, Gaetano and Luciano, Giuseppe Gaetano and Sheykhi, 
Ahmad",
    title = "{Slow-roll inflation and growth of perturbations in Kaniadakis 
modification of Friedmann cosmology}",
    eprint = "2307.04027",
    archivePrefix = "arXiv",
    primaryClass = "gr-qc",
    doi = "10.1140/epjc/s10052-023-12112-7",
    journal = "Eur. Phys. J. C",
    volume = "83",
    number = "10",
    pages = "936",
    year = "2023"
}

@article{Hernandez-Almada:2021aiw,
    author = "Hern{\'a}ndez-Almada, A. and Leon, Genly and Maga{\~n}a, Juan and Garc{\'\i}a-Aspeitia, Miguel A. and Motta, V. and Saridakis, Emmanuel N. and Yesmakhanova, Kuralay",
    title = "{Kaniadakis-holographic dark energy: observational constraints and global dynamics}",
    eprint = "2111.00558",
    archivePrefix = "arXiv",
    primaryClass = "astro-ph.CO",
    doi = "10.1093/mnras/stac255",
    journal = "Mon. Not. Roy. Astron. Soc.",
    volume = "511",
    number = "3",
    pages = "4147--4158",
    year = "2022"
}

@article{Luciano:2022eio,
    author = "Luciano, Giuseppe Gaetano",
    title = "{Gravity and Cosmology in Kaniadakis Statistics: Current Status and Future Challenges}",
    doi = "10.3390/e24121712",
    journal = "Entropy",
    volume = "24",
    number = "12",
    pages = "1712",
    year = "2022"
}

@article{Jacobson1995,
	author = {Ted Jacobson},
	title = {Thermodynamics of spacetime: The Einstein equation of state},
	journal = {Phys. Rev. Lett.},
	volume = {75},
	pages = {1260--1263},
	year = {1995},
	eprint = {gr-qc/9504004}
}

@article{Padmanabhan2005,
	author = {T. Padmanabhan},
	title = {Gravity and the thermodynamics of horizons},
	journal = {Phys. Rept.},
	volume = {406},
	pages = {49--125},
	year = {2005},
	eprint = {gr-qc/0311036}
}

@article{Starobinsky1980,
    author = "Starobinsky, Alexei A.",
    editor = "Khalatnikov, I. M. and Mineev, V. P.",
    title = "{A New Type of Isotropic Cosmological Models Without Singularity}",
    doi = "10.1016/0370-2693(80)90670-X",
    journal = "Phys. Lett. B",
    volume = "91",
    pages = "99--102",
    year = "1980"
}

@article{Liravi:2026xzm,
    author = "Liravi, Leila and Sheykhi, Ahmad",
    title = "{Slow-roll inflation in (dual) Kaniadakis cosmology}",
    eprint = "2605.18070",
    archivePrefix = "arXiv",
    primaryClass = "gr-qc",
    month = "5",
    year = "2026"
}

@book{Piattella2018lecture,
    author = "Piattella, Oliver F.",
    title = "{Lecture Notes in Cosmology}",
    eprint = "1803.00070",
    archivePrefix = "arXiv",
    primaryClass = "astro-ph.CO",
    doi = "10.1007/978-3-319-95570-4",
    isbn = "978-3-319-95569-8, 978-3-030-07060-1, 978-3-319-95570-4",
    publisher = "Springer",
    address = "Cham",
    year = "2018"
}

@article{BICEP:2021xfz,
    author = "Ade, P. A. R. and others",
    collaboration = "BICEP, Keck",
    title = "{Improved Constraints on Primordial Gravitational Waves using Planck, WMAP, and BICEP/Keck Observations through the 2018 Observing Season}",
    eprint = "2110.00483",
    archivePrefix = "arXiv",
    primaryClass = "astro-ph.CO",
    doi = "10.1103/PhysRevLett.127.151301",
    journal = "Phys. Rev. Lett.",
    volume = "127",
    number = "15",
    pages = "151301",
    year = "2021"
}

@article{keskin2022inflationary,
title={The inflationary era of the universe via Tsallis cosmology},
author={Keskin, AI},
journal={International Journal of Geometric Methods in Modern Physics},
volume={19},
number={01},
pages={2250005},
year={2022},
publisher={World Scientific}
}

@article{odintsov2017inflationary,
    author = "Odintsov, S. D. and Oikonomou, V. K.",
    title = "{Inflationary Dynamics with a Smooth Slow-Roll to Constant-Roll Era Transition}",
    eprint = "1703.02853",
    archivePrefix = "arXiv",
    primaryClass = "gr-qc",
    doi = "10.1088/1475-7516/2017/04/041",
    journal = "JCAP",
    volume = "04",
    pages = "041",
    year = "2017"
}

@article{nojiri2017modified,
    author = "Nojiri, S. and Odintsov, S. D. and Oikonomou, V. K.",
    title = "{Modified Gravity Theories on a Nutshell: Inflation, Bounce and Late-time Evolution}",
    eprint = "1705.11098",
    archivePrefix = "arXiv",
    primaryClass = "gr-qc",
    reportNumber = "PHYS.REPT.-692-(2017)-1-104, Phys.Rept. 692 (2017) 1-104",
    doi = "10.1016/j.physrep.2017.06.001",
    journal = "Phys. Rept.",
    volume = "692",
    pages = "1--104",
    year = "2017"
}

@article{hwang2005classical,
    author = "Hwang, Jai-chan and Noh, Hyerim",
    title = "{Classical evolution and quantum generation in generalized gravity theories including string corrections and tachyon: Unified analyses}",
    eprint = "gr-qc/0412126",
    archivePrefix = "arXiv",
    doi = "10.1103/PhysRevD.71.063536",
    journal = "Phys. Rev. D",
    volume = "71",
    pages = "063536",
    year = "2005"
}

@article{alhallak2023salvaging,
    author = "Alhallak, M. and Chamoun, N. and Eldaher, M. S.",
    title = "{Salvaging power-law inflation through warming}",
    eprint = "2212.04935",
    archivePrefix = "arXiv",
    primaryClass = "astro-ph.CO",
    doi = "10.1140/epjc/s10052-023-11667-9",
    journal = "Eur. Phys. J. C",
    volume = "83",
    number = "6",
    pages = "533",
    year = "2023"
}

@article{remmen2014many,
    author = "Remmen, Grant N. and Carroll, Sean M.",
    title = "{How Many $e$-Folds Should We Expect from High-Scale Inflation?}",
    eprint = "1405.5538",
    archivePrefix = "arXiv",
    primaryClass = "hep-th",
    reportNumber = "CALT-2014-138",
    doi = "10.1103/PhysRevD.90.063517",
    journal = "Phys. Rev. D",
    volume = "90",
    number = "6",
    pages = "063517",
    year = "2014"
}

@article{Salehi:2024kkj,
    author = "Salehi, A.",
    title = "{The analytical approach in testing the Kaniadakis cosmology}",
    doi = "10.1088/1361-6382/ad789f",
    journal = "Class. Quant. Grav.",
    volume = "41",
    number = "20",
    pages = "205012",
    year = "2024"
}

@article{ivanov2022analytic,
    author = "Ivanov, Vsevolod R. and Ketov, Sergei V. and Pozdeeva, Ekaterina O. and Vernov, Sergey Yu.",
    title = "{Analytic extensions of Starobinsky model of inflation}",
    eprint = "2111.09058",
    archivePrefix = "arXiv",
    primaryClass = "gr-qc",
    reportNumber = "IPMU21-0073",
    doi = "10.1088/1475-7516/2022/03/058",
    journal = "JCAP",
    volume = "03",
    number = "03",
    pages = "058",
    year = "2022"
}

@article{akrami2020planck,
   author = "Akrami, Y. and others",
    collaboration = "Planck",
    title = "{Planck 2018 results. X. Constraints on inflation}",
    eprint = "1807.06211",
    archivePrefix = "arXiv",
    primaryClass = "astro-ph.CO",
    doi = "10.1051/0004-6361/201833887",
    journal = "Astron. Astrophys.",
    volume = "641",
    pages = "A10",
    year = "2020"
}
\end{document}